\documentclass[fleqn,usenatbib]{mnras}

\usepackage{newtxtext,newtxmath}

\usepackage[T1]{fontenc}

\DeclareRobustCommand{\VAN}[3]{#2}
\let\VANthebibliography\thebibliography
\def\thebibliography{\DeclareRobustCommand{\VAN}[3]{##3}\VANthebibliography}

\usepackage{graphicx}	
\usepackage{amsmath}	

\newcommand{\nusth}{$\nu_{\rm{STH}}$}

\newcommand{\nualpha}{$\nu_{\alpha}$}
\newcommand{\nuc}{$\nu_{c}$}
\newcommand{\chir}{$\chi_r^2$}

\title[Universal model for dark matter halo density profiles]{Towards a universal model for the density profiles of dark matter haloes}

\author[S. T. Brown et al.]{
Shaun T. Brown,$^{1}$\thanks{E-mail: S.T.Brown@2018.ljmu.ac.uk}
Ian G. McCarthy,$^{1}$\thanks{E-mail: I.G.McCarthy@ljmu.ac.uk}
Sam G. Stafford,$^{1}$
Andreea S. Font$^{1}$
\\
$^{1}$Astrophysics Research Institute, Liverpool John Moores University, 146 Brownlow Hill, Liverpool L3 5RF \\
}

\date{Accepted XXX. Received YYY; in original form ZZZ}

\pubyear{2021}

\begin{document}
\label{firstpage}
\pagerange{\pageref{firstpage}--\pageref{lastpage}}
\maketitle

\begin{abstract}
It is well established from cosmological simulations that dark matter haloes are not precisely self-similar and an additional parameter, beyond their concentration, is required to accurately describe their spherically-averaged mass density profiles.  We present, for the first time, a model to consistently predict both halo concentration, $c$, and this additional `shape' parameter, $\alpha$, for a halo of given mass and redshift for a specified cosmology. Following recent studies, we recast the dependency on mass, redshift, and cosmology to a dependence on `peak height'.  We show that, when adopting the standard definition of peak height, which employs the so-called spherical top hat (STH) window function, the concentration--peak height relation has a strong residual dependence on cosmology (i.e., it is not uniquely determined by peak height), whereas the $\alpha$--peak height relation is approximately universal when employing the STH window function.  Given the freedom in the choice of window function, we explore a simple modification of the STH function, constraining its form so that it produces universal relations for concentration and $\alpha$ as a function of peak height using a large suite of cosmological simulations. It is found that universal relations for the two density profile parameters can indeed be derived and that these parameters are set by the linear power spectrum, $P(k)$, filtered on different scales. We show that the results of this work generalise to any (reasonable) combination of $P(k)$ and background expansion history, $H(z)$, resulting in accurate predictions of the density profiles of dark matter haloes for a wide range of cosmologies. 
\end{abstract}

\begin{keywords}
cosmology: dark matter, theory -- methods: numerical
\end{keywords}



\section{Introduction}

The mass density profile of dark matter (DM) haloes is a key prediction of the current concordance $\Lambda$CDM cosmology.  The density distribution has been shown to depend both on the mass of a halo and redshift, with the precise dependencies being set by the cosmological parameter values that specify the initial conditions and expansion rate of the Universe \citep[e.g.][]{Frenk_1988}. 

It has been shown in many previous studies that the density profiles of DM haloes can be reasonably well approximated by an NFW profile \citep{NFW1,NFW2}:
\begin{equation}
    \rho (r) =\frac{ \rho_0}{(r/r_s)(1+r/r_s)^2} \ \ \ , 
\end{equation}
\noindent where $r_s$ is the scale radius, often quoted as a concentration ${c=R/r_s}$ (where $R$ is the halo radius, usually defined using a spherical overdensity definition), and $\rho_0$ is the normalisation, which can be constrained by the total mass of the halo.  A key prediction of this formalism is that the structure of DM haloes, as a function of mass, requires a single free parameter, the scale radius or concentration. Consequently, many empirical and analytic models have been developed to try to accurately predict the concentration of haloes as a function of mass, redshift, and cosmological parameters \citep[e.g.][]{Bullock_2001,Eke_2001,Prada_2012,Ludlow_2014,Diemer_2015,Correa}.

Although it is common to describe the density profiles of DM haloes through a scale radius (i.e. a single parameter), it has been demonstrated that DM haloes are not perfectly self-similar and that a second parameter (other than concentration) is required to accurately describe the density profiles.  This is true for both individual and stacked density profiles \citep[e.g.][]{Gao_2008,Navarro_2010}.  The Einasto profile \citep{einasto1965} has been shown to better reproduce the density profiles observed in high-resolution simulations:
\begin{equation} \label{eqn:Einasto}
    \ln (\rho (r)/\rho_{-2} )= -\frac{2}{\alpha} \bigg [ \bigg ( \frac{r}{r_{-2}} \bigg )^{\alpha}-1 \bigg ] \ \ \ .
\end{equation}
Here $r_{-2}$ is again a scale radius, defined to be the radius where the logarithmic slope $d\ln{\rho}/d\ln{r}$, is equal to $-2$, and is therefore equivalent to $r_s$ used in the NFW parameterisation.  The parameter $\alpha$ is commonly referred to as the `shape' parameter and describes how quickly the slope of the density profile varies as a function of radius. For $\alpha \approx 0.18$, the Einasto profile closely resembles an NFW form over radii typically sampled in cosmological simulations. 

As shown in \cite{Gao_2008}, the parameter $\alpha$ exhibits a clear dependence on both halo mass and redshift, and also has a dependence on the underlying cosmology as later demonstrated by \cite{Ludlow_2016}.  Therefore, both $c$ and $\alpha$ depend on mass, redshift, and cosmology, motivating a model that can consistently predict both parameters for a general cosmology.  Compared to the halo concentration, the shape parameter has received relatively little attention in the literature and as such there does not yet exist a model aimed at predicting $\alpha$ for a general cosmology, only empirical models that predict the $\alpha$--$M$ relation for a specific cosmology \citep[e.g.][]{Duffy_08,Ludlow_2013,Dutton_2014}. Note that a significant number of models infer the concentration of haloes from simulations adopting a fixed shape parameter when fitting to the density profiles, which can lead to biased estimates of concentration (as the two parameters are not independent).\footnote{In principle the radius where the logarithmic slope is $-2$ can be directly estimated independent of $\alpha$, and the concentration can be defined with this radius. However, in practice this is very rarely done and when fitting the measured density profile over a wide radial range $c$ and $\alpha$ are not independent.} Clearly $\alpha$ and $c$ should be modelled in a consistent way to be able to reliably predict the density profiles of DM haloes.

In this paper we aim to link changes to both the initial density fluctuations, i.e., the linear power spectrum $P(k)$, and the background expansion, $H(z)$, to the resulting density profiles of DM haloes and to quantify the dependence on halo mass and redshift.  Ideally, predictions for both $c$ and $\alpha$ should fit into a consistent and physically-motivated theoretical framework.  Following recent work, we recast the dependencies on halo mass and redshift into a single dependence on `peak height', a quantity which characterises the amplitude of density fluctuations with respect to some critical threshold for collapse (see Section~\ref{section:peak height definition} for a general definition).  The use of peak height is well motivated by the spherical collapse model \citep{Gunn&Gott_1972} and plays an important role in the successful (extended) Press-Schechter formalism \citep{Press-Schechter}.  Previous simulation work has shown that peak height correlates very strongly, though not perfectly, with both $c$ and $\alpha$ \citep[e.g.][]{Prada_2012,Gao_2008}. In this work we re-examine the definition of peak height and explore the freedom therein in order to derive accurate universal relations (i.e., applicable for wide ranges of cosmological parameters) for $c$ and $\alpha$.  Specifically, we exploit the freedom in the form of the window function that is used to filter the linear power spectrum when computing the peak height.  We will show that the standard window function, the so-called spherical top hat function, is not the optimal choice for predicting the density profile parameters and that a relatively simple modification thereof results in a substantially improved model. 

The paper is organised as follows. In Section~\ref{section:technical_section} we discuss the technical details of the simulations and how they are processed, particularly focusing on how the density profiles are stacked and fitted to obtain values and errors for $c$ and $\alpha$. In Section~\ref{section:peak height definition} we discuss the definition and key properties of peak height and how it is calculated for a given cosmology. In Section~\ref{section:peak height relations} we  present how the two density parameters, $c$ and $\alpha$, vary with peak height for the cosmologies using the standard definition. In Section~\ref{section:window function} we motivate the use of an alternative window function when defining peak height and quantitatively determine an optimal choice so that both $c$ and $\alpha$ are universally described by this new definition of peak height.  We additionally develop and present the model to predict halo concentration and shape parameter for a general cosmology. In Section~\ref{Section:EdS test} we test our model using additional cosmologies with very different background expansion rates to those used to calibrate our model. Finally, in Section~\ref{Section:summary} we conclude and summarise our results.

\begin{figure*}
    \centering
    \includegraphics[width=2\columnwidth]{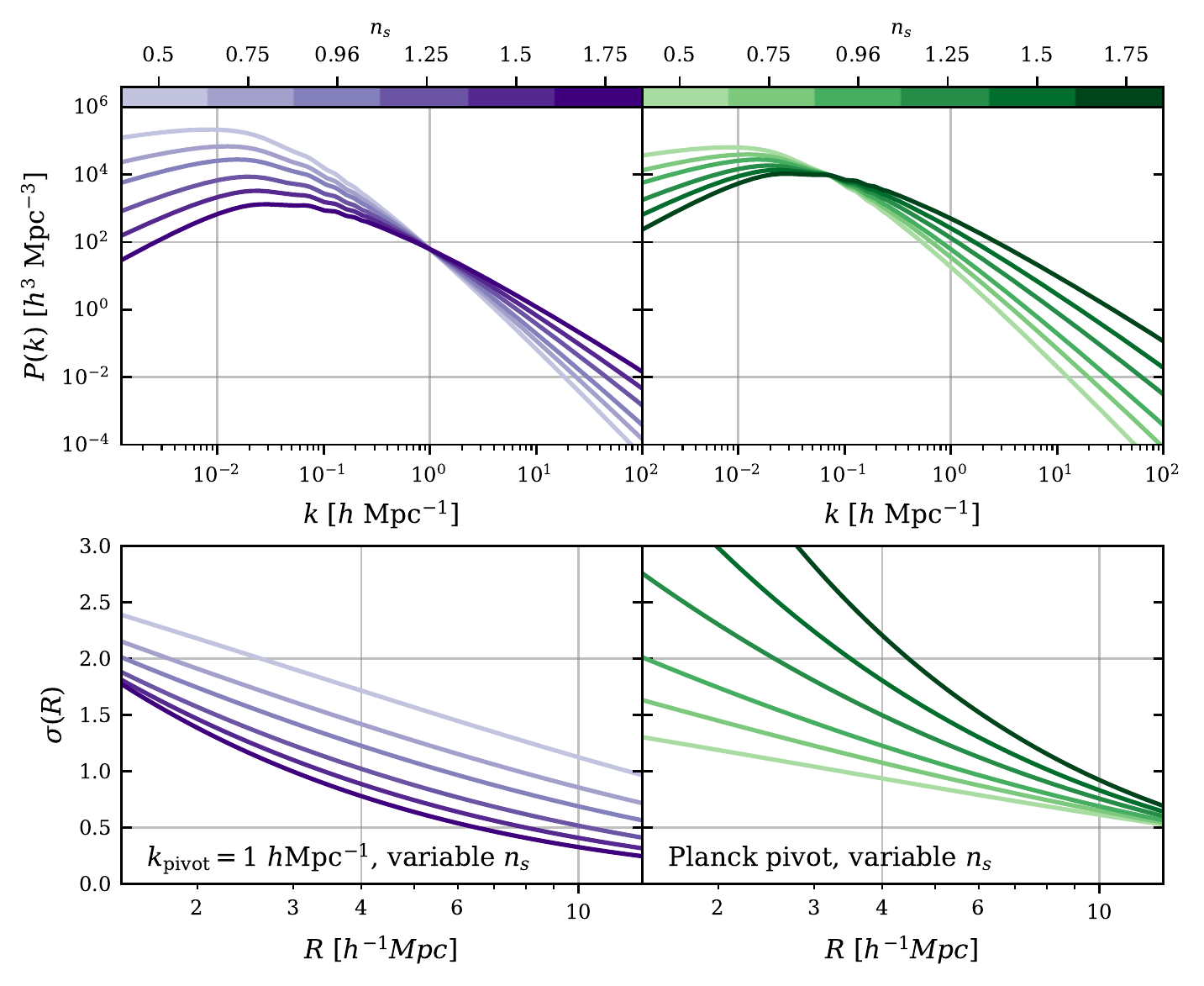}
    \caption{\textit{Top:} The $z=0$ linear power spectra for the various cosmologies studied in this work.  \textit{Bottom:} The rms density fluctuations in spheres of radius $R$. The left and right panels represent the two different suites (introduced in \citealt{brown2020}), which use different pivot points for the linear power spectrum (see label in bottom left). For each suite (or pivot point), the primordial spectral index, $n_s$ is systematically varied from $0.5$ to $1.75$ with $n_s=0.96$ being the best-fit WMAP 9-yr value. The different shades represent different values of $n_s$, see colour bar.}
    \label{fig:power_spectra}
\end{figure*}

\section{Simulation and analysis details} \label{section:technical_section}

In this section we present the various cosmologies studied as well as the technical details of the simulations used in this work.  We also describe how the density profiles are calculated and fitted to determine values for halo concentration, $c$, and shape parameter, $\alpha$.

\subsection{Cosmologies}
\label{section:cosmologies}

In this work we primarily study a subset of the cosmologies first presented in \cite{brown2020}, particularly examining the two suites closest to our own universe.  We discuss here briefly these different cosmologies. For a more in depth description of how these cosmologies were chosen we refer the reader to Section~2.3 of \cite{brown2020}.

The cosmologies presented in \cite{brown2020} were chosen to systematically study the effects of changes to both the amplitude and shape (i.e. slope) of the linear power spectrum at different $k$-scales on the internal properties of dark matter haloes, such as the density and velocity dispersion profiles. The amplitude and shape were changed by using a combination of free parameters in the $\Lambda$CDM model: the primordial amplitude, $A_s$, the primordial spectral index, $n_s$, which directly affects the slope of the linear power spectrum and $k_{\rm{pivot}}$, which is the $k$-scale used for normalising the linear power spectra. The cosmologies used in the present work are split into two suites, the `Planck pivot' and `$k_{\rm{pivot}}=1h$ Mpc$^{-1}$' suites. For each suite the primordial spectral index, $n_s$, is systematically varied from $0.5$ to $1.75$  with a fixed $A_s$ and $k_{\rm{pivot}}$. The $n_s=0.96$ case represents the best-fit WMAP 9-yr results and is therefore a close match to to what we believe is our own Universe. We present the values of $A_{s}$, $n_s$, $k_{\rm{pivot}}$ and $\sigma_8$ for these different cosmologies in Table.~\ref{simulation_table}. These cosmologies share the same best-fit WMAP 9-yr background expansion (Friedmann) parameters: $h=0.7$, $\Omega_{\rm{m}}=0.2793$, $\Omega_{\rm{b}}=0.0463$ and $\Omega_{\Lambda}=0.7207$ \citep{hinshaw2013}.

In Fig.~\ref{fig:power_spectra} we present the $z=0$ linear power spectra (top panels) for the $k_{\rm{pivot}}=1h$ Mpc$^{-1}$ (left panels, purple lines) and Planck pivot (right panels, green lines) suites. The different shades represent the different values of $n_s$, as shown by the colour bars above each column. The two different pivot points can clearly be seen at $k_{\rm{pivot}}=1h$ Mpc$^{-1}$ and $k_{\rm{pivot}}=0.05$ Mpc$^{-1}$ (Planck pivot point), allowing for the power spectra to be normalised at different physical scales.  Additionally plotted in the bottom panels is the root mean square (rms) density fluctuations as a function of Lagrangian radius.  Note that $\sigma (R)$, which is formally defined in Eqn.~(\ref{eqn:peak_height}) below, correlates strongly with the expected amount of structure and abundance of haloes at different scales and masses, with larger mass haloes corresponding to larger Lagrangian radii, and vice-versa. We discuss these quantities further in Section~\ref{section:peak height definition}. 

As can been seen in Fig.~\ref{fig:power_spectra}, the cosmologies studied in this work represent a wide range of different shapes and amplitudes to the linear power spectrum, which in turn results in a diverse amount of expected structure, as described through $\sigma (R)$. This results in a sample of haloes with widely different evolutions and formation histories, offering a broad context in which to study the cosmological dependence of the density profiles of DM haloes.

\begin{table}
\caption{Summary of the various cosmological parameters for the majority of simulations presented in this work. The main two parameters varied are $n_s$ and $A_s$. Along with $k_{\rm{pivot}}$, they specify the initial power spectrum for a $\Lambda$CDM cosmology. We also present values of $\sigma_8$ for these cosmologies. All cosmologies have the same background expansion: $h=0.7$, $\Omega_{\rm{m}}=0.2793$, $\Omega_{\rm{b}}=0.0463$ and $\Omega_{\Lambda}=0.7207$.}
\label{simulation_table}
\begin{tabular}{lllllllll}
\hline
Simulation suite                                 & $n_s$  & $A_s$ [$10^{-9}$]                 & $k_{\rm{pivot}} [h$~Mpc$^{-1}]$ & $\sigma_8$  \\
\hline
WMAP9 best fit                  & $0.96$ & $2.392$   & $2.86 \times 10^{-3}$         & $0.801$               \\

\textit{Planck} pivot                    & $0.5$  & $2.103 $                     & $7.14 \times 10^{-2}$         & $0.687$             \\
\textit{Planck} pivot                   & $0.75$ & $2.103 $                      & $7.14 \times 10^{-2}$                           & $0.743$           \\
\textit{Planck} pivot                   & $1.25$ & $2.103 $                      & $7.14 \times 10^{-2}$                            & $0.904$            \\
\textit{Planck} pivot                    & $1.5$  & $2.103 $                      & $7.14 \times 10^{-2}$                            & $1.016$             \\
\textit{Planck} pivot                   & $1.75$ & $2.103 $                      & $7.14 \times 10^{-2}$                            & $1.154$      \\   
$k_{\rm{pivot}}=1~h$ Mpc$^{-1}$  & $0.5$  & $1.892$                      & $1.00$                           & $1.261$          \\
$k_{\rm{pivot}}=1~h$ Mpc$^{-1}$ & $0.75$ & $1.892 $                      & $1.00$                            & $0.980$           \\
$k_{\rm{pivot}}=1~h$ Mpc$^{-1}$ & $1.25$ & $1.892 $                      & $1.00$                            & $0.616$            \\
$k_{\rm{pivot}}=1~h$ Mpc$^{-1}$  & $1.5$  & $1.892$                      & $1.00$                            & $0.498$            \\
$k_{\rm{pivot}}=1~h$ Mpc$^{-1}$ & $1.75$ & $1.892 $                      & $1.00$                            & $0.407$          \\

\hline
\end{tabular}
\end{table}

\subsection{Simulation details}

The simulations studied in this work are virtually identical to those presented in \cite{brown2020}; the only difference being that all cosmologies from the original work have been re-run with a box twice the size (but with the same mass resolution), resulting in a factor of $8$ increase in volume.  This was done to increase the number of haloes in each simulation, allowing larger mass haloes to be studied as well as improve the statistics at all masses.  Other than the box size, the technical details are the same as for the simulations presented in \cite{brown2020}, which we describe below.

The linear power spectra are generated using the Boltzmann code \texttt{CAMB} \citep{camb}. Initial particle positions and velocities are calculated using a modified version of \text{N-GenIC}\footnote{The publicly available version of this code can be found at \url{https://github.com/sbird/S-GenIC}.} \citep{Gadget_2} at a starting redshift of $z=127$.  The initial conditions include second-order Lagrangian perturbation theory corrections and identical phases are adopted for all simulations. The collisionless, or `DM-only', N-body simulations have been run with a modified version of the \texttt{Gadget-3} code \citep{Gadget_2,bahamas}. The simulations have been run with a comoving periodic volume of size $400$ $h^{-1}\textrm{Mpc}$ on a side with $1024^3$ particles. For a WMAP 9-yr background cosmology \citep{hinshaw2013}, as used for the majority of cosmologies in this work, this corresponds to a particle mass of $4.62 \times 10^{9}h^{-1}$ M$_{\odot}$. The gravitational softening is fixed to $4 h^{- 1}$kpc (in physical coordinates for $z \leq 3$ and in co-moving at higher redshifts).

All haloes are identified with the \texttt{SUBFIND} algorithm \citep{Subfind}. In this work we present the spherically averaged density profiles of DM haloes, using the most bound particle of the central halo as the halo centre. The central halo is defined as the largest (sub)halo in the friend-of-friends (FOF) group. We calculate the density using all particles within the given spherical shell, whether they are identified as belonging to a subhalo or not. In principle the density of the smooth component with substructure removed can also be calculated \citep[e.g.][]{Fielder_2020}. In general, the halo finder is primarily used to initially identify the FOF group, provide the location of the centre of potential and calculate bulk properties such as halo mass and radius (for a given definition).

\subsection{Fitting density profiles}

As already stated, the goal of this work is to accurately study and model both $c$ and $\alpha$ for a wide range of cosmologies. It is therefore essential that the simulation data are processed in an appropriate way to obtain reliable and robust measures of $c$ and $\alpha$ with their associated errors.

Throughout this work we will exclusively fit `stacked' density profiles, described as follows. The spherically-averaged density profile of individual haloes are calculated using $32$ logarithmically spaced bins over the radial range $10^{-2.5}<r/R_{\rm{200c}}<0.7$, where $R_{\rm{200c}}$ is a measure of the halo size (see Section~\ref{section:peak height definition} for definition). The stacked profile is then calculated as the median density in each radial bin from all the haloes in the stack of a given halo mass bin. The values of $c$ and $\alpha$ are then calculated by fitting the stacked density profiles with an Einasto profile (see Eqn.~(\ref{eqn:Einasto})) such that the following figure of merit, $\psi$, is minimised:
\begin{equation}
    \psi^2=\sum_i[\log \rho_i(r) -\log \rho_{\rm{Einasto}}(r)]^2 \ \ \ ,
\end{equation}
here $\rho_i(r)$ is the density profile from the simulation and $\rho_{\rm{Einasto}}(r)$ is the Einasto profile for a given set of parameters.  To estimate the errors on $c$ and $\alpha$ we use bootstrap resampling.  Specifically, $1,000$ different realisations of the stacked profile are generated by randomly sampling (with repetition) haloes within the mass bin (or stack) using the same number of haloes. Hence, the number of haloes in the stacked profile depends strongly on mass, redshift and the cosmology. The values of $c$ and $\alpha$ are then estimated as the median of the resulting distribution with the lower and upper errors calculated as the $16^{\rm{th}}$ and $84^{\rm{th}}$ percentiles respectively, equivalent to a $1\sigma$ uncertainty for a Gaussian distribution.

Although the stacked density profiles are calculated over a relatively large radial range only a subset of these radial bins are actually used when fitting to obtain values for $c$ and $\alpha$. In this work we fit over the radial range $r_{\rm{conv}}<r<0.7 R_{\rm{200c}}$, where $r_{\rm{conv}}$ is the convergence radius and dictates the minimum radius before numerical uncertainties affect the density profiles, which is discussed below. 

The maximum radius of $0.7 R_{\rm{200c}}$ was chosen to avoid the very outer parts of a halo that are potentially not in equilibrium and do not follow the NFW or Einasto forms particularly well \citep[e.g.][]{Ludlow_2010}. The minimum radius we adopt is the so-called convergence radius, $r_{\rm{conv}}$, that specifies the radius at which the density profile is subject to numerical effects. Specifically, at small radii two-body interactions lead to a resolution dependent density core. The convergence radius can therefore be expressed as a ratio of the timescale of two-body interactions and the Hubble time via:
\begin{equation} \label{eqn:convergence}
    \frac{r_{\rm{conv}}}{R_{\rm{200c}}}=4 \bigg (\frac{\kappa_{\rm{P03}} \ln N_c}{\sqrt{N_{\rm{200c}}N_c}} \bigg )^{2/3} \ \ \ ,
\end{equation}
where $N_c$ is the number of particles below the convergence radius and $\kappa_{\rm{P03}}$ is the ratio of the collisional relaxation time and the age of the universe \citep{Power_convergence,Ludlow_convergence}. Larger values of $\kappa_{\rm{P03}}$ represent a more conservative convergence criterion. \citet{Power_convergence} propose that a value of $\kappa_{\rm{P03}}=0.6$ leads to convergence for individual haloes, while \citet{Ludlow_convergence} find a similar, though smaller, value of $\kappa = 0.18$ for the convergence of stacked density profiles. However, in this work we find a larger value is needed to provide unbiased estimates for $c$ and $\alpha$. We find that $\kappa_{\rm{P03}}=7$\footnote{Using $\kappa_{\rm{P03}}=7$ means that for the cosmologies with the highest halo contractions, specifically the $n_s=1.75$ cosmology in the Planck pivot suite, have $r_{\rm{conv}} \sim r_{-2}$. Though the scale radii are well resolved for the majority of haloes.} provides reliable results and this value has also been suggested for improved convergence by \cite{Navarro_2010}. It is found that, as well as a clear numerical core forming at the centre of haloes as documented in these works, there also occurs a slight enhancement in the density at larger radii of $r \approx r_{\rm{conv}}$ (for $\kappa_{\rm{P03}}=0.6$), as is expected in order to conserve halo mass. Using $\kappa_{\rm{P03}}=0.6$ does avoid fitting to the density profile where there is a significant suppression in the density but typically does not avoid fitting to the region exhibiting an enhancement in density. Although this enhancement in density is relatively small, typically at most $\approx 5 \%$, the difference can propagate through to $\approx 20\%$ systemic differences when determining the best-fit values of $c$ and $\alpha$. This appears to only be a significant issue when fitting the density profiles with a free shape parameter, due to the increased versatility of the fit. If a fixed shape parameter is used, either by explicitly fixing $\alpha$ or using a fitting formula without an equivalent `shape' term, such as an NFW profile, then the determination of $c$ is only mildly affected by the systematic differences in the inner density profile. It is likely that $\kappa_{\rm{P03}}=7$ is somewhat overly conservative, but it does ensure that there are no systematic errors associated with either the numerical core or the aforementioned enhancement in density. In this work we explicitly solve Eqn.~(\ref{eqn:convergence}), with $\kappa_{\rm{P03}}=7$, for all individual haloes within a stack and use the median $r_{\rm{conv}}$ when fitting the stacked density profiles. For a more detailed discussion of the convergence radius, including derivations and alternative forms to Eqn.~(\ref{eqn:convergence}), we refer the reader to \citet{Power_convergence} and \citet{Ludlow_convergence}.

\begin{figure}
    \centering
    \includegraphics[width=\columnwidth]{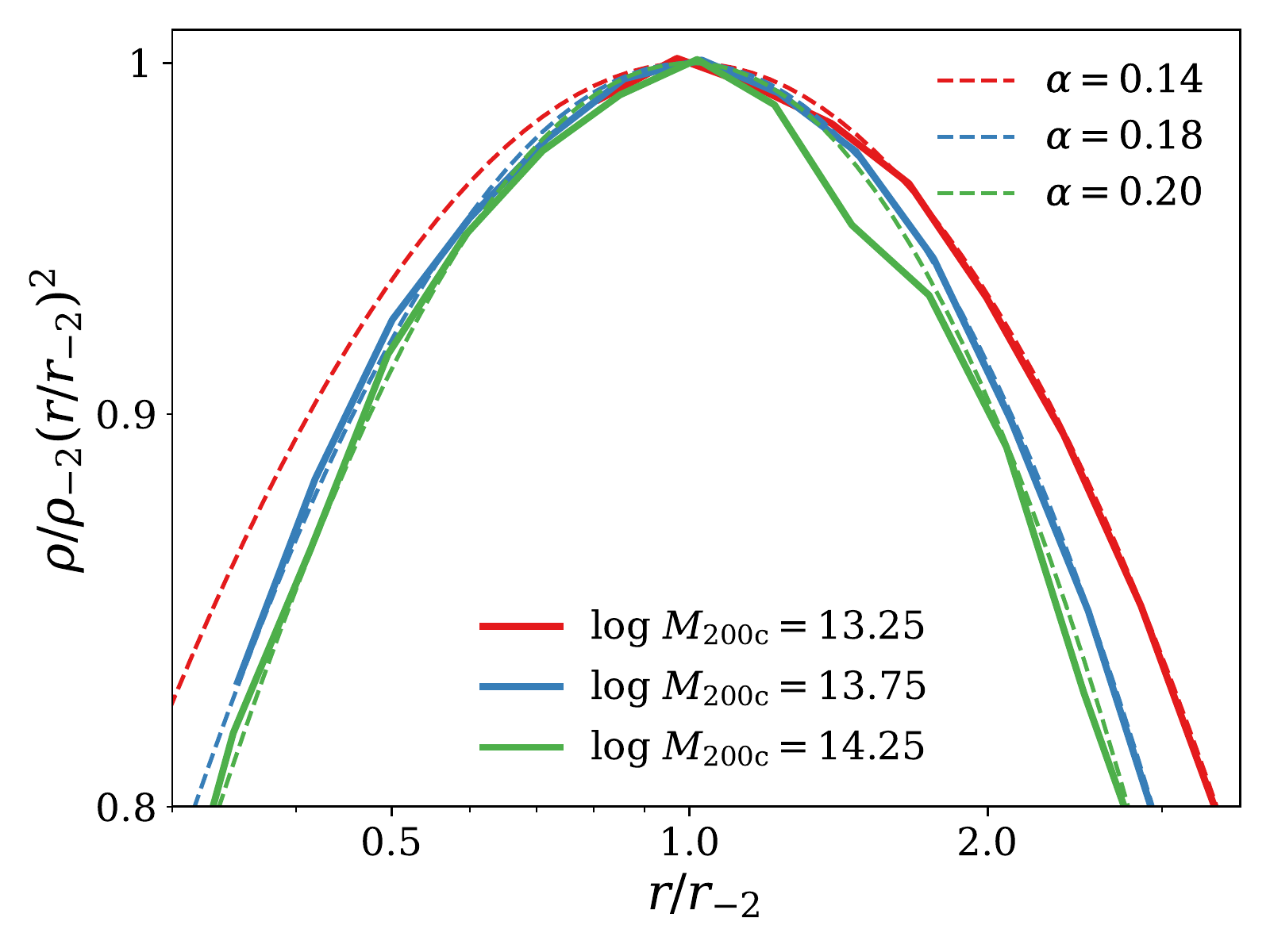}
    \caption{Stacked density profiles for a range of masses (see legend) at $z=0$ for a WMAP 9-yr cosmology. The $\log$ $M_{\rm{200c}} = 13.25$, $13.75$ and $14.25$ mass bins constitute stacks of $5,713$, $1,318$ and $211$ haloes, respectively. The profiles are only plotted up to their convergence radius (Eqn.~(\ref{eqn:convergence})) which varies strongly with halo mass. For each density profile, $r_{-2}$ and $\rho_{-2}$ are estimated non-parametrically from the logarithmic slope. The density profiles are normalised by their respective scale radii, $r_{-2}$, and plotted as $\rho r^2$ to reduce the dynamic range. Normalising the radial coordinate in this way removes the dependence on concentration.  As can be seen, there is a clear halo mass dependence to the normalised density profiles, demonstrating that the density profiles are not self-similar and that an additional `shape' parameter is required to fully describe them.  Plotted as dashed lines are Einasto profiles that approximately follow the simulated density profiles.  In these units, the Einasto profile has only one free parameter, $\alpha$ (see legend). }
    \label{fig:density_profile}
\end{figure}

As discussed above, we focus on studying stacked density profiles to derive both the $c$ and $\alpha$ mass relations. For most of the cosmologies studied we use logarithmically-spaced mass bins of width $0.3$ dex for haloes with at least $5,000$ particles.  For our simulations this results in a minimum mass of $M_{\rm{200c}}=2.31 \times 10^{13}$ $h^{-1} \rm{M_{\odot}}$. The only exception to this is the most extreme cosmology considered in this work, using a primordial spectral index of $n_s=1.75$ with a Planck pivot point, where we apply a cut of $10,000$ particles (see Appendix~\ref{section:res_study} for details).  We also only consider mass bins with at least 100 haloes. The maximum halo mass considered therefore depends strongly on the halo mass function and varies as a function of cosmology and redshift. 

As well as imposing a cut on both the number of particles within a halo as well as the number of haloes within a stack, we also apply a relaxation cut to discount haloes that have been significantly affected by ongoing or recent major mergers. Specifically, we only consider haloes with a normalised offset of the centre of mass (CoM) to centre of potential (CoP) of $| \mathbf{x}_{\rm{CoP}}- \mathbf{x}_{\rm{CoM}} | /R_{\rm{200c}} <0.07$. This relaxation criterion is similar to that presented in \citet{Neto}, who applied the same cut to the CoP and CoM offset but with additional criteria based on the relative mass of substructure and the virial ratio. We find that simple cut on the CoM to CoP offset is sufficient to remove severely unrelaxed haloes and gives unbiased estimates for $c$ and $\alpha$. Additionally, applying the extra criteria proposed in \citet{Neto} does not significantly change the inferred $c$ and $\alpha$ values, a similar conclusion to that found in \citet{Duffy_08}.

In Fig.~\ref{fig:density_profile} we present examples of the density profiles that are fit in this work.  Here we have plotted the profiles for a range of masses at $z=0$ for a WMAP 9-yr cosmology, normalised by $r_{-2}$ and $\rho_{-2}$. The radius $r_{-2}$ and is taken to be where the logarithmic slope equals $-2$ and is found by directly interpolating the logarithmic slope of the stacked density profiles.\footnote{To reliably estimate the slope of the profiles we use a Savitzky–Golay filter, with a window length of $3$ and a second order polynomial.} This allows for $r_{-2}$ and $\rho_{-2}$ to be determined empirically from the density profiles directly, without any assumptions about the parametric form the overall density profile may take.  Plotting in these units, i.e. $r/r_{-2}$ and $\rho/\rho_{-2}$, removes the dependence on halo concentration. If the density profiles were perfectly self-similar they should be indistinguishable when plotted in this manner. However, as shown in Fig.~\ref{fig:density_profile} the density profiles (solid lines) do not follow the same radial dependence as each other, with higher (lower) masses resulting in profiles where the $\rho r^2$ profile varies more quickly (slowly) with radius. This difference demonstrates the need for an additional parameter other than concentration; i.e., the shape parameter. When plotted in these units, i.e. $r_{-2}$ and $\rho_{-2}$, the Einasto profile has only one free parameter, $\alpha$ (see Eqn.~(\ref{eqn:Einasto})). Additionally plotted as dashed line in Fig.~\ref{fig:density_profile} is a number of Einasto profiles with values for $\alpha$ chosen by eye to approximately follow to observed density profiles. It can be seen that the role of $\alpha$ is to control how quickly $\rho r^2$ varies with radius.

If one uses the definition that the scale radius is where the logarithmic slope is $-2$ then the concentration of haloes can, in principle, be determined separately from the shape parameter and any assumptions about the density profile, as we have done above. However, practically it is often more reliable to determine $c$ and $\alpha$ by fitting directly to the density profiles, as is done in this work (see Section~\ref{section:technical_section}) and many others. When determining $c$ and $\alpha$ in this way they are no longer independent, and there will be a certain amount of degeneracy between the two parameters. Additionally the value of $c$ inferred by fitting to the density profile will depend on the assumed density profile, including, for an Einasto profile, if $\alpha$ is allowed to be free or not.

Fig.~\ref{fig:density_profile} also demonstrates the main sources of errors when fitting to stacked profiles at different mass scales. Specifically the number of haloes within the stack and the limited radial range fit over. For higher mass bins (green line) the main limiting factor is the relatively small number of haloes within the mass range, resulting in a  somewhat noisy density profile with relatively large fluctuations.  For lower masses (red line) there are many more systems resulting in a much smoother density profile, however, there is a significantly reduced radial range over which the profiles can be reliably measured due to the low number of particles in each halo and hence a larger convergence radius.

In Appendix~\ref{section:res_study} we present a resolution study to check the convergence of both our simulations and analysis and we motivate further some of our selection choices, such as only considering haloes with at least $5,000$ particles.

\begin{figure*}
    \centering
    \includegraphics[width=2\columnwidth]{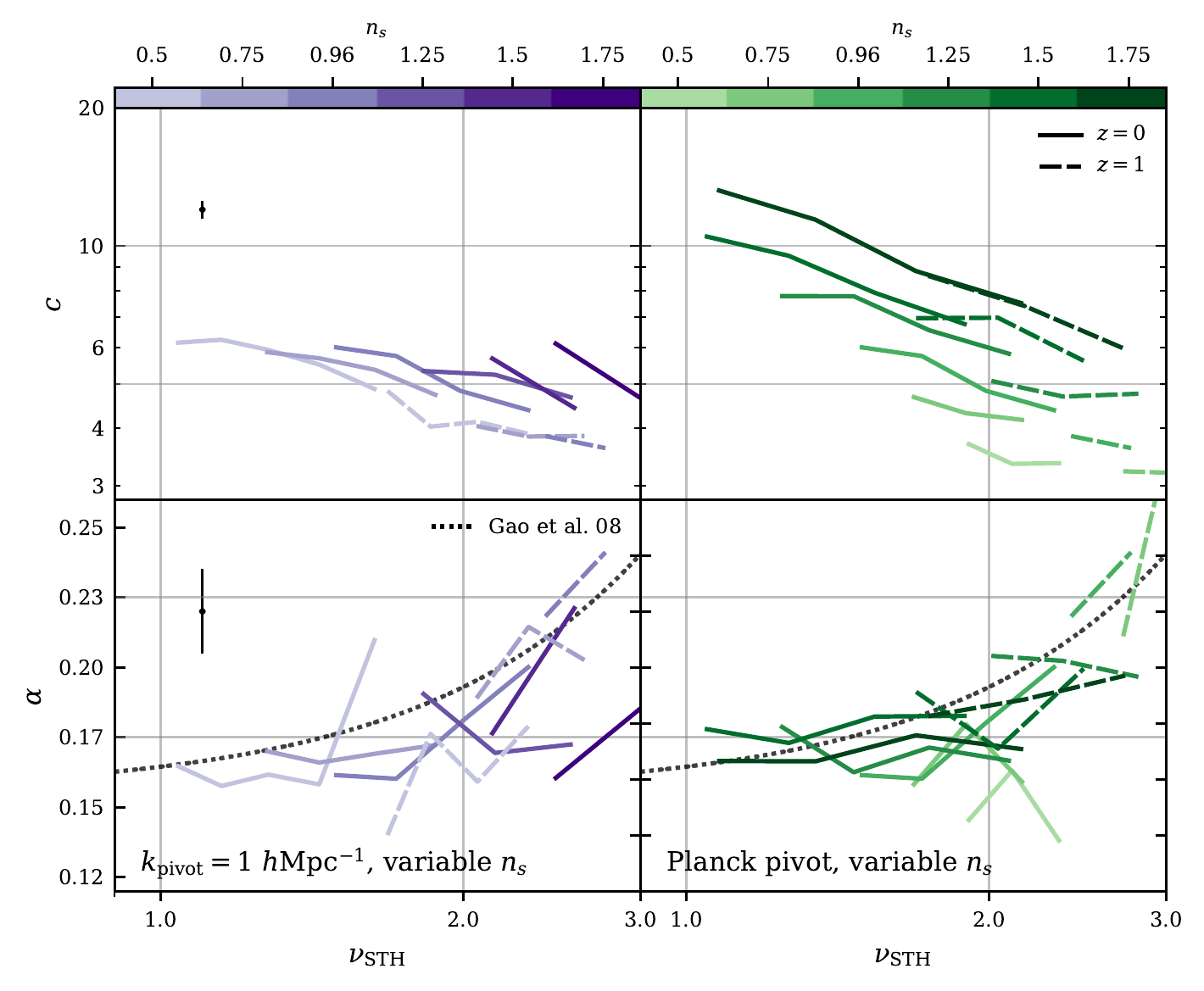}
    \caption{Halo concentration, $c$, and shape parameter, $\alpha$, as a function of peak height computed using the standard STH window function, $\nu_{\rm{STH}}$. The main cosmologies used in this work are plotted, with the specific cosmology given by the colour and shade of the lines.  In purple is the suite using $k_{\rm{pivot}}=1 h$ Mpc$^{-1}$ while in green the suite using a Planck pivot, the shade of the colour indicates the primordial spectral index, as shown by the colour bar above each panel. The line style indicates the redshift for the given cosmology (see legend). Additionally plotted as the black dotted line is the relation between $\alpha$ and $\nu_{\rm{STH}}$ presented in \protect\cite{Gao_2008}. For each data point shown here there is an associated error which we have not shown to improve the clarity and legibility of the plot. For reference the mean error is shown as the black error bar in the top left of each panel. In general it can be seen that there is both a cosmology and redshift dependence to the $c$--\nusth~relation. However, there is not a clear dependence on either redshift or cosmology for the $\alpha$--\nusth~relation, with all data points following approximately a single function.}
    \label{fig:peak_hight_relation}
\end{figure*}

\section{Peak height definition} \label{section:peak height definition}

Throughout this work we discuss how the density profiles of haloes vary as a function of peak height for a range of cosmologies. Here we discuss the definition of peak height and highlight some of the free aspects of the formalism where certain choices, or assumptions, have to be made; primarily the halo mass definition and window function used.

Peak height, $\nu$, is traditionally defined as,
\begin{equation} \label{eqn:peak_height_1}
    \nu(M,z)=\frac{\delta_c}{\sigma(M,z)} \ \ \ ,
\end{equation}
where $\delta_c$ is the critical density for collapse, as predicted by the spherical collapse model,\footnote{We take $\delta_c=1.68$ and ignore the mild cosmology dependence. We also do not consider any additional dependence on mass present in ellipsoidal collapse models.}  and $\sigma$ is the rms overdensity associated with the halo, and is calculated from the linear power spectrum via:
\begin{equation}
    \label{eqn:peak_height}
    \sigma^2(R,z)=\frac{1}{2 \pi^2} \int\displaylimits^{\infty}_{0} k^2P(k,z) |W(kR)|^2 dk \ \ \ .
\end{equation}
In the above, $P(k,z)$ is the linear power spectrum and $W(kR)$ is the window function (sometimes referred to as the filter function). $R$ is the so-called Lagrangian radius defined as
\begin{equation} \label{eqn:lagrangian_rad}
     R^3= \frac{M}{4/3 \pi \rho_{\rm{m},0}} \ \ \ ,
\end{equation}
where $M$ is the halo mass (e.g., the virial mass or that corresponding to some other spherical overdensity) and $\rho_{m,0}$ is the mean background density of the universe today.

The redshift evolution of the linear power spectrum can be written as
\begin{equation}
    P(k,z)=D^2(z)P(k,z=0) \ \ \ ,
\end{equation}
where $D(z)$ is the linear growth factor, which can be calculated from the background expansion, i.e. $H(z)$, and is normalised to unity at the current epoch.  The redshift evolution of peak height can therefore also be expressed in terms of the growth function, where
\begin{equation}
    \nu(M,z)=\nu(M,z=0)/D(z) \ \ \ .
\end{equation}
Hence for a given cosmology, with $P(k)$ and $H(z)$ specified, the peak height of a halo can be straightforwardly calculated from the above equations.

There are a few key aspects in the above equations that are open to interpretation and therefore certain choices must be made. The first of these is how the mass of a halo is defined.  It is common to define the mass as an overdensity with respect to either the critical or mean density of the universe at a given redshift. The mass, and in turn radius, of the halo is defined to obey the following,
\begin{equation}
    \Delta \rho_{\rm{c/m}}=M_{\Delta \rm{c/ m}}/(4/3 \pi R_{\Delta \rm{c/ m}}^3) \ \ \ ,
\end{equation}
where $\Delta$ is the adopted overdensity. We will use the notation specified in this equation to identify the given mass definition, identifying both the overdensity parameter (from the number in the subscript) and reference density (with either denoting c or m for using the critical or mean density, respectively). Common choices are $M_{\rm{200c/m}}$ and $M_{\rm{500c/m}}$. The definition with respect to the higher overdensity and smaller radius, $M_{\rm{500c/m}}$, is typically used for galaxy clusters (since X-ray observations typically probe the hot gas within this radius), while $M_{\rm{200c/m}}$ is often used when examining the properties of dark matter haloes, of all sizes, in cosmological simulations. It has been shown that large scale properties of DM halos, such as the abundance at a given mass or the position of the splashback radius, correlate more strongly with a $M_{\rm{200m}}$ definition \citep[e.g.][]{Tinker_2008,Diemer_2014,Diemer_2020}.  On the other hand, internal properties, such as halo concentration, tend to correlate more strongly with an $M_{\rm{200c}}$ mass definition \citep[e.g.][]{Diemer_2015}. Why different properties of dark matter haloes seem to prefer a mass definition with respect to either the critical or mean density is unclear and is an open question in the field. In this work we are focused on studying and developing a model for the density profiles of DM haloes and we use a mass definition of $M_{\rm{200c}}$.   We leave the exploration of alternative mass definitions for future work, though we do briefly discuss this possibility in the context of our results in Section~\ref{section:optimal window func}.

The second aspect of peak height formalism for which there is freedom is in the choice of the window function, $W(kR)$, which is the main focus of this paper.  It has become common place in the literature that $W(kR)$ is chosen so that it represents a spherical top hat (STH) function in configuration (real) space. With this choice, the window function takes the following form:
\begin{equation} \label{eqn:sth}
    W_{\rm{STH}}(kR)=\frac{3}{(kR)^3}[\sin(kR)-kR\cos(kR)] \ \ \ .
\end{equation}

This choice of window function provides a very obvious and clear interpretation of Eqn.~(\ref{eqn:peak_height}); it represents the rms overdensity averaged over a sphere of radius R for the given linear power spectrum.  This choice also makes it clear how to compare to and use the spherical collapse model, which considers the evolution of a top hat perturbation in an otherwise homogeneous expanding universe. However, the spherical collapse model does not offer a complete picture of how real haloes assemble, particularly ignoring the hierarchical growth that is at the heart of our current cosmological paradigm.  It is therefore not clear that this is the correct choice of window function for such a cosmology and potentially a different choice of window function would better represent (or correlate with) the growth and structure of haloes in a cold DM dominated universe. 

Throughout this paper we will use a subscript to identify the window function used to calculate peak height. For instance \nusth~refers to peak height values calculated using the standard spherical top hat window function. We reserve the use of $\nu$ \textit{without} a subscript when discussing peak height in a general sense with, in principle, any choice of window function, as we have above.

\section{Peak height relations} \label{section:peak height relations}

Before proceeding to study if $c$ and $\alpha$ can be better modelled by an alternative window function, it is worth studying how these density parameters vary as a function of peak height using the standard STH definition.

In Fig.~\ref{fig:peak_hight_relation} we present how $c$ and $\alpha$ vary as a function of \nusth~for the main cosmologies studied in this work (see Section~\ref{section:cosmologies}), at $z=0$ and $z=1$. We have chosen not to plot individual errors (the mean error is plotted in black at the top left of each row) to improve the readability of the plot, but note that not all values here are equally reliable with some data points having significantly larger fractional errors than others. In general, larger values of \nusth~result in fewer haloes within the mass bin and therefore larger uncertainties. The relation between $\alpha$ and \nusth~proposed by \citet{Gao_2008} is shown as the dotted black line, specifically $\alpha=0.0095\nu_{\rm{STH}}^2+0.155$. 

Focusing initially on the concentration of the haloes at $z=0$ (top panels, solid lines), we see that there is a clear cosmological dependence to the $c$--\nusth~relation. This is particularly clear for the Planck pivot suite where the different cosmologies are significantly stratified. Additionally, for a given cosmology, \nusth~does not appear to completely describe the redshift evolution. This is most easily seen for the $k_{\rm{pivot}}=1$ $h$Mpc$^{-1}$ suite where haloes at $z=1$ (dashes lines) have significantly lower concentrations at a fixed \nusth~than at $z=0$ (solid lines).

The dependence of $\alpha$ on \nusth~(bottom panels) is much closer to universal than for $c$. In general both the cosmological and redshift dependences appear to be well described by \nusth~with no obvious trend of a certain cosmology or redshift lying distinct from the main distribution, as is observed for $c$. Our $\alpha$--\nusth~relation matches reasonably well that previously proposed by \cite{Gao_2008}. 

From these results there is clearly room for improving the universality of the relation between $c$ and peak height, which may potentially be achieved by altering the window function (away from the standard STH case) in the peak height definition.  On the other hand, the standard definition of peak height already correlates very well with $\alpha$ in a way that is apparently independent of cosmology and redshift.  This suggests that the STH window function is already close to optimal for $\alpha$.  Taken together, these results suggest that $c$ and $\alpha$ favour separate and distinct window functions. Indeed, this is what we find in the next section.

\begin{figure*}
    \centering
    \includegraphics[width=\columnwidth]{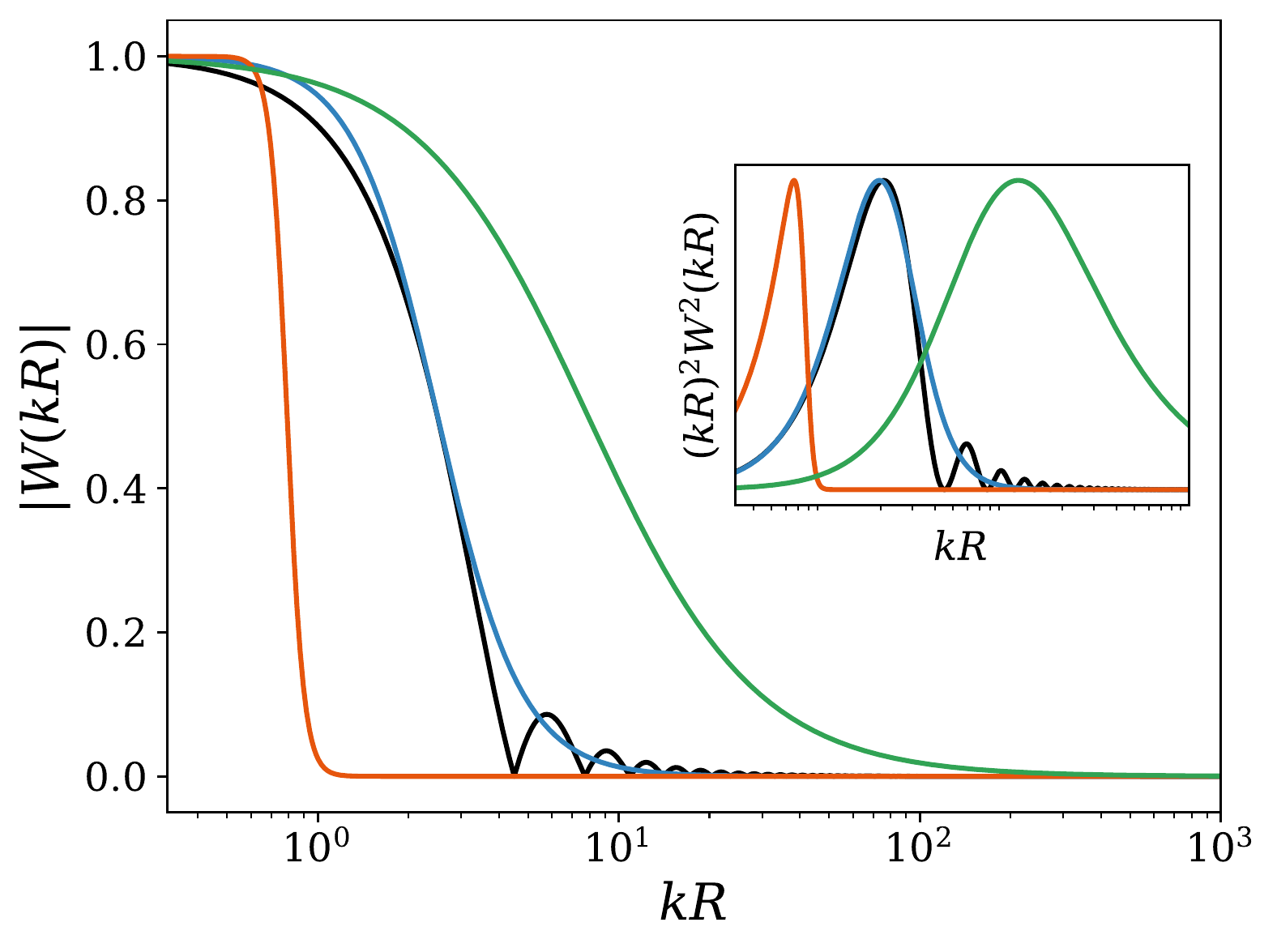}
    \includegraphics[width=\columnwidth]{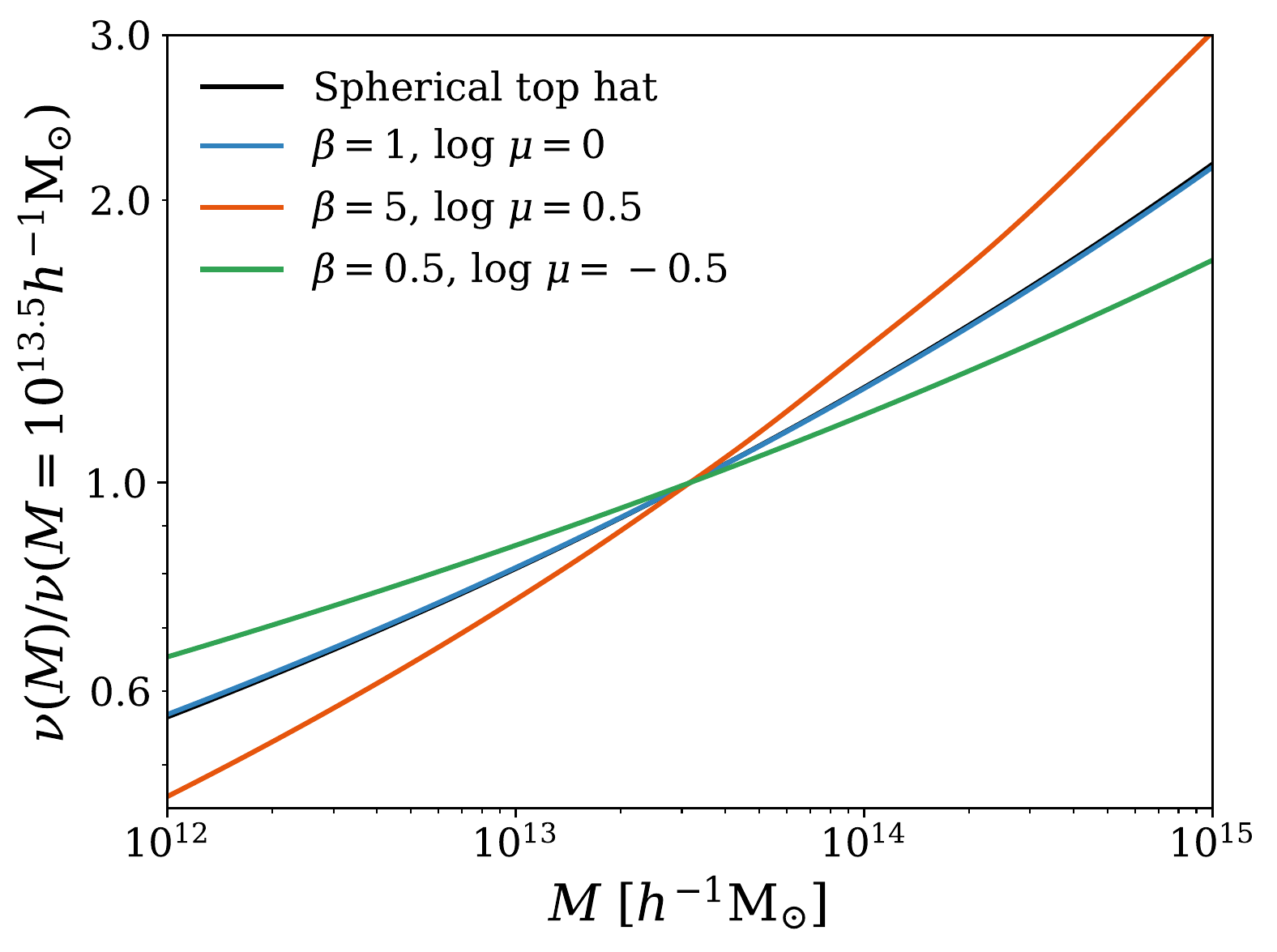}
    
    \caption{{\it Left:} The smooth $k$-space window function (Eqn.~(\ref{eqn:smooth_k})) for a few combinations of $\mu$ and $\beta$ (see legend). Qualitatively, $\mu$ changes the scale at which the transition occurs, with smaller values of $\mu$ resulting in the transition occurring at higher values of $kR$ (corresponding to smaller physical scales), while $\beta$ controls how quickly the transition from $W=1$ to $W=0$ occurs. Plotted for reference is the standard STH window function (shown in black). Additionally plotted in the inset panel is $(kR)^2 W^2(kR)$ for the same window functions in the main plot, with each curve has been normalised by its global maximum. No units have been plotted for the inset, as the purpose of the figure is to demonstrate that $(kR)^2 W^2(kR)$, for all window functions studied here, exhibits a clear peak, with the location of that peak depending on both $\mu$ and $\beta$. {\it Right:} The resulting relation between peak height and mass, normalised at $M = 10^{13.5} h^{-1} M_{\odot}$ for the WMAP-9 yr cosmology. The STH (black lines) and smooth $k$-space window function with $\mu=1$ and $\beta=1$ (blue lines) follow a very similar $\nu (M)$ relation.}
    \label{fig:window_func}
\end{figure*}

\section{Optimal window functions} \label{section:window function}

In a $\Lambda$CDM universe, where the initial density fluctuations are assumed to be small and Gaussian in nature, the initial density field can be, statistically, described by the power spectrum alone.  The subsequent gravitational evolution depends only on these initial conditions and the background expansion history, given a theory for gravity. Hence, the averaged internal structure of haloes as a function of mass depends only on the linear power spectrum, $P(k)$, and Hubble expansion, $H(z)$, which together form a given cosmology. 

The aim of the present study is to determine how the averaged structure of haloes depends on $P(k)$ and $H(z)$ in a quantitative fashion.  A promising theoretical framework to use is the Press-Schechter formalism \citep{Press-Schechter}. In the Press-Schechter formalism the abundance of haloes is predicted to be a universal function of peak height. This has motivated previous studies to also correlate halo properties with peak height, as it is expected that peak height will account for a significant part of the cosmological dependence.  However, the abundance of haloes is only approximately universal and numerical simulations have shown that there is a clear redshift and cosmology dependence when using the standard Press-Schechter formalism, i.e. a STH window function \citep[e.g.][]{Tinker_2008}. There have been a number of suggestions for how to improve this model, with a notable extension being the use of an alternative window function. For instance, \cite{Leo_2018} found that using a smooth $k$-space filter can accurately model the abundance of haloes in cosmologies with truncated power spectra (e.g., due to non-standard inflation scenarios). By allowing the window function to vary, we are able to study what aspects of $P(k)$ are most important for setting the density parameters, where it is found that both density parameters are approximately set by the amplitude of $P(k)$ at an associated k-scale (see Section~\ref{section:optimal window func}).

As demonstrated in the previous section, the density parameters, particularly $c$, are clearly not universal as a function of peak height when using a STH window function. Similar to how \cite{Leo_2018} found the abundance of haloes can be closer to universal with an alternative window function, it is possible that a different window function will result in $c$--$\nu$ and $\alpha$--$\nu$ relations that are universal; i.e., do not depend on redshift or cosmology.

It is worth considering if the use of an alternative window function is consistent with some of the key results from the literature as well as the features already observed in Section~\ref{section:peak height relations}. 

The first result we consider is from \cite{Diemer_2015}, who study the concentration of haloes in scale free cosmologies, that being a cosmology with a power law linear power spectrum and an Einstein de Sitter background expansion (i.e., $\Omega_{\Lambda}=0$ and $\Omega_{m}=1$). They find that for a single cosmology the redshift evolution closely follows a single function of \nusth. However, the particular relation between $c$ and \nusth~exhibits a clear dependence on the choice of the slope of the linear power spectrum (see Fig.~3 of \citealt{Diemer_2015}). Their interpretation was that it is the effective slope of the linear power spectrum at an associated $k$-scale that affects the $c$--$\nu$ relation and led \citet{Diemer_2015} to develop a model that incorporates an effective slope in addition to peak height in order to better predict halo concentration.  However, this is not the only interpretation of their results, it is also consistent with the possibility of using a different window function. For a power law linear power spectrum, $P(k)=A k^n$, calculating $\sigma$ (see Eqn.~(\ref{eqn:peak_height})) is somewhat simplified:
\begin{equation}
    \sigma ^2 (R,z) = \frac{A}{2 \pi ^2} D^2(z) R^{-(n+3)} \int\limits ^{\infty} _{0} x^{n+2} |W(x)|^2 dx.
\end{equation}
Therefore, $\sigma (R,z) \propto D^2(z)R^{-3-n}$. It is clear that for these cosmologies the window function only plays a role in the normalisation of the peak height. Therefore, for a given cosmology, any choice of window function would preserve $\nu$--$c$ being redshift independent. However, when comparing different cosmologies, i.e. different values of $n$, the window function and its effect on the normalisation will play a role, as can be seen by the $x^{n+2}$ term within the integrand. It is therefore likely that the window function could be chosen appropriately so that the normalisation between different values of $n$ would result in a single $c$--\nusth~relation independent of $n$, or the normalisation of the power spectrum. 

Another key result that should be accounted for, or preserved, is the redshift evolution of $c$ and $\alpha$ in a standard $\Lambda$CDM cosmology. It is well established that the redshift evolution of $c$ is not perfectly described by \nusth, while \nusth~offers a good description of the redshift evolution of $\alpha$. Therefore developing a model to predict $c$ and $\alpha$ must predict this general behaviour. For a $\Lambda$CDM cosmology the linear power spectrum is no longer a power law, meaning that the window function contributes in a more complex way to the peak height of different mass haloes than simply by a different normalisation. Therefore, changing the window function from the standard STH case can potentially change the relationship between peak height and mass in such a way as to resolve the discrepancy in the redshift evolution of $c$; meanwhile if the window function remains relatively close to the STH case then the redshift evolution of $\alpha$ can be preserved. 

One potential limitation of any model that links the density profiles of DM haloes to only $P(k)$ and $H(z)$, as is the goal of this work, is that the density profiles of individual haloes cannot be predicted. Due to the statistically-averaged nature of the power spectrum, our model predicts averaged quantities for $c$ and $\alpha$. As such, in this paper we focus exclusively on modelling the averaged density parameters as a function of mass and redshift. However, it seems inevitable that the density profiles of individual haloes depends, in detail, on the initial overdensity in the linear power spectrum with which they are associated. Therefore, differences in these overdensities would correspond to differences in individual halo density profiles. It is therefore likely that the theoretical framework presented in this paper could be extended to describe the expected scatter in $c$ and $\alpha$ for a fixed mass, however, this is beyond the scope of this paper.

An alternative approach to that presented in this work is to identify an appropriate mediator that correlates strongly with the density profiles of haloes. For example, it is common to attribute halo concentration with the formation history of the halo \citep[e.g.][]{NFW2,Wechsler_2006,Ludlow_2014}. This therefore offers a natural explanation for the general mass dependence, with smaller haloes forming earlier and resulting in higher concentrations, as well as the observed scatter in $c$ at fixed mass corresponding to an equivalent scatter in formation time. However, formation history is not a fundamental property and depends on the given cosmology. Hence, to make a prediction for a given cosmology (i.e. $P(k)$ and $H(z)$), some theoretical framework is required to predict the formation history as a function of mass, redshift and cosmology. Extended Press-Schechter theory \citep[e.g][]{Lacy&Cole_93} is one such theoretical framework that aims to predict the distribution of formation histories. A prescription for the link between formation history and halo concentration can therefore be used alongside such a theoretical framework to predict the distribution of expected halo concentrations \citep[e.g.][]{Benson_2019}.

\subsection{Smooth \texorpdfstring{$k$}-space window function} \label{section:window function definitions}

Assuming a correct choice of window function exists there is no obvious way to derive, from first principles, the form that it should take. As such we use a more heuristic approach by using a versatile parameterisation for the window function that maintains key properties that are expected to be present for a realistic window function.

We use the \textit{smooth $k$-space} window function originally proposed in \cite{Leo_2018} to study how the $c$--$\nu$ relation changes for various choices of window function. The smooth $k$-space window function is defined as the following,
\begin{equation}
    \label{eqn:smooth_k}
    W_{\rm{smooth}}(kR)=\frac{1}{1+(\mu kR/2.50)^{3.12 \beta}} \ \ \ .
\end{equation}
The smooth $k$-space window function behaves very similarly to a step function (in fourier-space), with $\beta$ determining how quickly the transition from $0$ to $1$ occurs\footnote{As $\beta \rightarrow \infty$ the function reduces exactly to a step function.} and $\mu$ the scale at which the transition happens. When defining Eqn.~(\ref{eqn:smooth_k}) we have normalised the free parameters to resemble closely the standard STH window function when $\mu$ and $\beta$ are unity. This is done so that for ${\mu=1}$, ${\beta=1}$ the scale where $W(kR)=0.5$ and the first derivative at that scale match the standard STH window function. This results in the factors of $2.50$ and $3.12$. In practice, this means that the results for a choice of $\mu=1$ and $\beta=1$ when using the smooth $k$-space window function will resemble closely the spherical top hat case, allowing for an easier interpretation of these parameters compared to the standard definition for peak height. In the left hand panel of Fig.~\ref{fig:window_func} we show the smooth $k$-space function for a few combinations of $\mu$ and $\beta$, and discuss the inset panel later in Section~\ref{section:optimal window func}. The standard STH function (see Eqn.~(\ref{eqn:sth})) is plotted for comparison. As can be seen in Fig.~\ref{fig:window_func}, the smooth $k$-space window function can closely resemble the standard STH filter (by construction at $\mu=1$, $\beta=1$). One feature that cannot be replicated is the series of `wiggles' at high values of $kR$, however these do not contribute significantly to the peak height calculation.  

In the right hand panel of Fig.~\ref{fig:window_func} we show the resulting peak height values as a function of mass for the standard WMAP 9-yr cosmology at $z=0$, normalised by the peak height at $M=10^{13.5}$ $h^{-1}$ M$_{\odot}$. This demonstrates that the relation between mass and peak height depends intimately on the choice of window function. When the window function is changed significantly so does the relationship between mass and peak height, which therefore propagates through to changes in the $c$ and $\alpha$ peak height relations. Note that $\nu(M)$ is almost indistinguishable when using the STH window function or the smooth $k$-space filter with $\mu=1$ and $\beta=1$. 

\begin{figure*}
    \centering
    \includegraphics[width=\columnwidth]{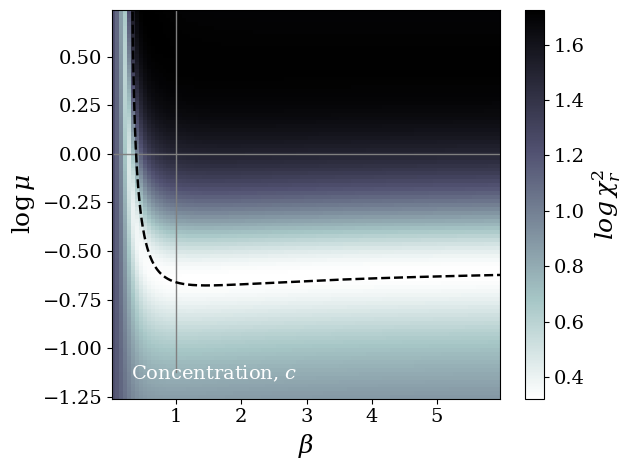}
    \includegraphics[width=\columnwidth]{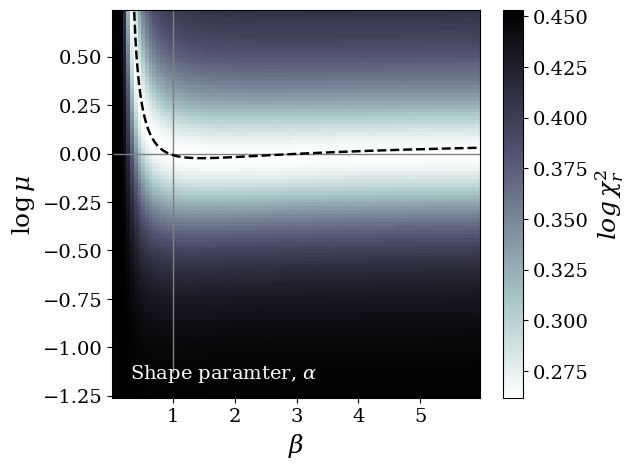}
    \caption{Parameter space demonstrating how \chir~varies with the free parameters of the smooth $k$-space window function, $\mu$ and $\beta$, for the two density parameters $c$ ({\it left}) and $\alpha$ ({\it right}). \chir~is used to quantitatively determine how close to universal the resulting peak height relations are, with smaller values of \chir~corresponding to more universal relations. To first order the value of peak height is set by the amplitude of the linear power spectrum at an associated $k$-scale, with that scale depending on the given window function. The key property is where $(kR)^2 W^2(kR)$ is a maximum, as described through the parameter $\kappa$, see Eqn.(\ref{eqn:k-scale}). Plotted as dashed black lines are contours of constant $\kappa$ (see Eqn.~(\ref{eqn:const_kappa})), with $\kappa=9$ and $\kappa=2$ for $c$ and $\alpha$, respectively. These contours follow very closely the observed degeneracies between $\mu$ and $\beta$. }
    \label{fig:smooth-kspace}
\end{figure*}

\begin{figure}
    \centering
    \includegraphics[width=\columnwidth]{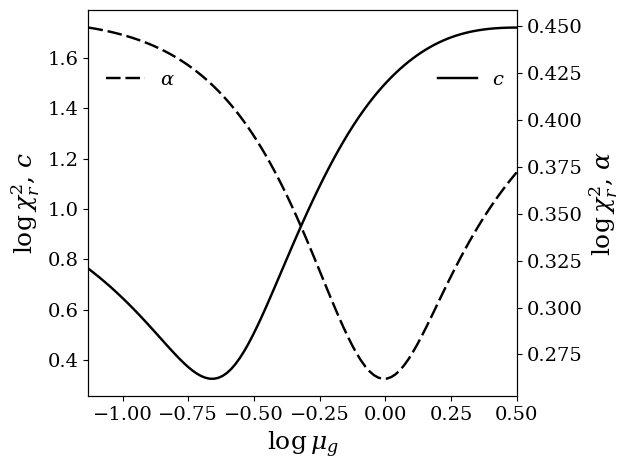}
    
    \caption{Variation of \chir~as a function of $\mu_g$ for the generalised spherical top-hat window function (see Eqn.~(\ref{eqn:sth_generalised})) for the two density parameters $c$ (solid line) and $\alpha$ (dashed line). \chir~is used to quantitatively determine how close to universal the resulting peak height relations are, with smaller values of \chir~corresponding to more universal relations. These distributions exhibit clear minima at $\log \mu_g=-0.67$ and $\log \mu_g=-0.01$ for $c$ and $\alpha$, respectively.  Note that $\log \mu_g=0$ ($\mu_g =1$) corresponds to the standard spherical top-hat window function.}
    \label{fig:optimal_gSTH}
\end{figure}

\subsection{Quantitatively determining universality} \label{section:determining universality}

As mentioned previously, the aim of this paper is to determine if, with an appropriate window function, the density profiles are universal with peak height. Practically this means that both $c$ and $\alpha$ follow a single function of peak height, $\nu$, for any $P(k)$ at any redshift. The cosmologies studied previously in \cite{brown2020} (see Section \ref{section:cosmologies}) offer a wide range of different linear power spectra that is ideal to constrain the optimal window function(s).  

To determine the optimal window function and constrain the associated parameters we require an appropriate figure of merit that quantitatively describes how close to a single function, and hence how universal, the resulting $c$ and $\alpha$--peak height relations are.  We choose to fit a second order polynomial, in log space, that minimises the $\chi^2$ error. The fitting formula is specifically
\begin{equation} \label{eqn:second-polynomial}
    \log(y)=a_2 \log(\nu)^2+a_1 \log(\nu)+a_0 \ \ \ .
\end{equation}
Here $y$ represents the parameter being constrained, either concentration, $c$, or the shape parameter, $\alpha$. The $\chi^2$ value for a given choice of $a_{0,1,2}$ and window function is calculated as,
\begin{equation} \label{eqn:chi2}
    \chi^2=\sum_i \frac{(y-y_i)^2}{\sigma_i^2} \ \ \ ,
\end{equation}
with the sum over all data points. $y$ is the given prediction for a choice of $a_{0,1,2}$, $y_i$ and $\sigma_i$ represent the value and error of the given data point. $a_{0,1,2}$ are then chosen to minimise Eqn.~(\ref{eqn:chi2}) for the given window function. Throughout the paper we will quote the reduced $\chi^2$ error, $\chi^2_{r}=\chi^2/DoF$, with the number of degrees of freedom (DoF) remaining constant.

In this work we want to study whether an appropriate choice of window function can lead to a universal relation between the two density parameters, $c$ and $\alpha$, and peak height, $\nu$. Therefore, the exact form that the $c$--$\nu$ or $\alpha$--$\nu$ relations take is of secondary importance compared to it obeying a single function for all cosmologies and redshifts studied. As such, we do not consider $a_{0,1,2}$ free parameters of the model, as they are only used to quantitatively determine `universality'. 

Using a second order polynomial in log space offers a fitting function that is versatile enough to describe the data without introducing higher order terms that could lead to overfitting. Ideally, a non-parametric method that does not impose a functional form on the $c$--$\nu$ and $\alpha$--$\nu$ relations would be used. One such method would be to minimise the Spearman rank correlation coefficient, which makes no assumptions about the functional form of the underlying data (other than the relation being monotonic). However, it is important to incorporate the associated errors for the density parameters, and it is unclear how to reliably include these in such a ranked statistic. 

\begin{figure*}
    \centering
    \includegraphics[width=\columnwidth]{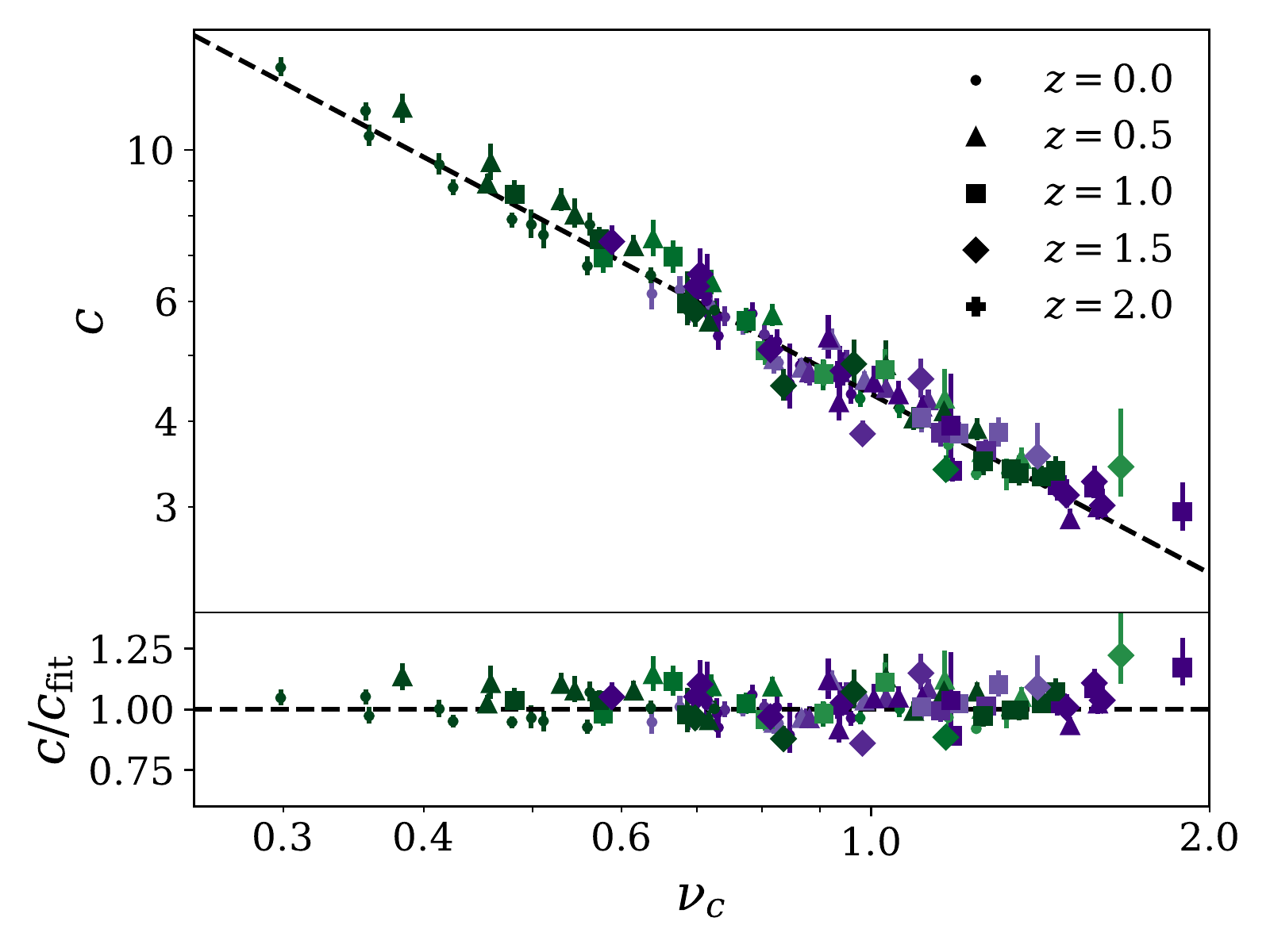}
    \includegraphics[width=\columnwidth]{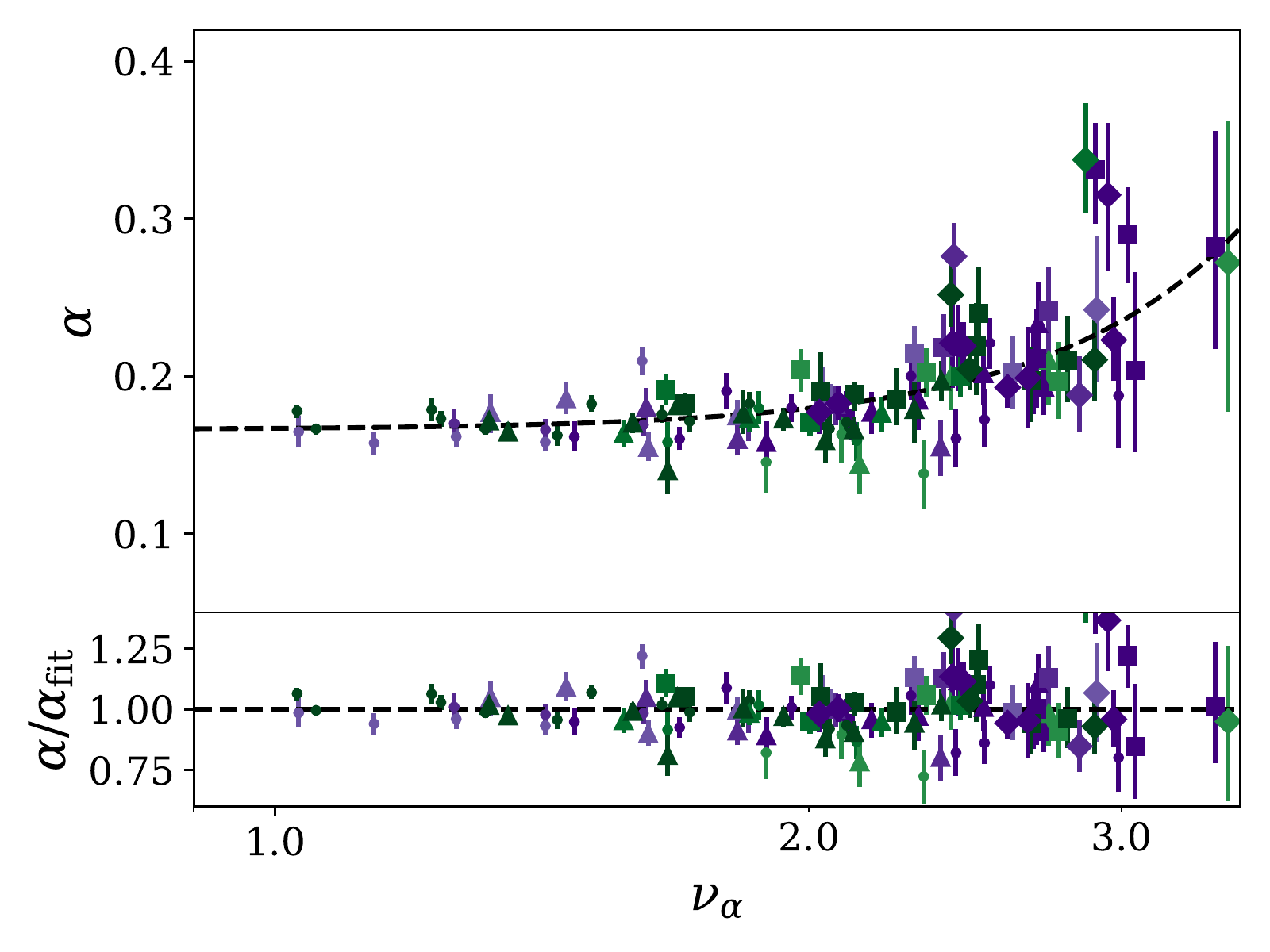}
    \caption{Resulting $c$--\nuc, left panel, and $\alpha$--\nualpha, right panel, relations for the optimal choice of window function. The data points presented are all those used to constrain the window function. The colour indicates the given cosmology, matching those in Fig.~\ref{fig:peak_hight_relation}, while the marker style corresponds to the redshift (see legend).  The black dashed lines represent the empirical relations used for our model to predict $c$ and $\alpha$ for a general cosmology. The fractional difference from the data and the empirical fits are shown in the bottom panels, for both $c$ and $\alpha$.}
    \label{fig:optimal_peak_height}
\end{figure*}

\subsection{An optimal window function} \label{section:optimal window func}

In Fig.~\ref{fig:smooth-kspace} we show how \chir~changes as a function of $\mu$ and $\beta$ when using a smooth $k$-space window function (Eqn.~(\ref{eqn:smooth_k})). The optimal window function is constrained separately for $c$ and $\alpha$ (left and right panels, respectively). Firstly, it is clear that there does indeed appear to be a choice of window function that results in a universal $c$--$\nu$ and $\alpha$--$\nu$ relation with minimum values of \chir$\approx 2$ for both $c$ and $\alpha$.  This is more clearly shown in Fig.~\ref{fig:optimal_peak_height}, which presents the resulting $c$--$\nu$ and $\alpha$--$\nu$ relations for an optimal choice of window function parameters (we discuss this in detail in the next subsection). 

Although the minimal values of \chir~are comparable for $c$ and $\alpha$, there are significant differences in the range of \chir~values. This does not appear to be a reflection of the smooth $k$-space filter better describing one parameter over the other, but rather features that are intrinsic to the data, independent of the choice of window function. The primary reason for this difference is that $c$ varies over a much larger dynamic range than $\alpha$, meaning that relatively small changes in peak heights result in large differences to how universal the $c$--$\nu$ relation is, while $\alpha$ is much less sensitive to these changes.

Focusing initially on the parameter space for the smooth $k$-space window function constrained by halo concentration (left panel of Fig.~\ref{fig:smooth-kspace}), it is clear that there are strong degeneracies in determining the optimal values of $\mu$ and $\beta$. However, a choice of parameters close to the standard spherical top hat case ($\mu=1$, $\beta=1$) is clearly disfavoured, as expected from the earlier discussion (see Section~\ref{section:peak height relations}). For $\beta \gtrsim 1$, the value of $\mu$ is relatively well determined and the optimal values appear to be independent of $\beta$, favouring a value of $\log \mu \approx -0.7$. However, in this region there is little constraint on $\beta$ with all values sampled performing similarly well. It is clear that a key factor in determining the concentration is the $k$-scale of the transition of the window function, how `quickly' this transition occurs is of secondary importance. There also exist degeneracies in the region $\beta \lesssim 1$. Here the form of the degeneracy is much more complicated than for $\beta \gtrsim 1$, exhibiting a nontrivial dependence on $\mu$ and $\beta$. We discuss the origin and form of this degeneracy shortly.

Examining the constraints on $\mu$ and $\beta$ when optimising for the shape parameter $\alpha$ (right panel of Fig.~\ref{fig:smooth-kspace}), we see the same general behaviour as for concentration. The overall shape of the degeneracy in the parameter space is almost identical, except with it being translated to larger values of $\mu$ from what is found for $c$. Interestingly, the (approximate) STH window function ($\mu=1$, $\beta=1$) lies almost perfectly on the observed degeneracy and is therefore close to an optimal choice of parameters. Again we observe for $\beta \gtrsim 1$ that there is no constraint on $\beta$, but $\mu$ is relatively well constrained. The optimal value in this region is $\mu \approx 1$ as opposed to $\log \mu \approx -0.7$, as was observed for halo concentration. The optimal window function appears to be somewhat at odds with the results of \citet{Ludlow_2016} who found that the $\alpha$--$\nu_{\rm{STH}}$ relation was not universal. There is no obvious explanation for this, but may be linked to the very different cosmologies studied in their work, specifically scale free cosmologies with EdS background expansions.

To further understand the observed degeneracies between $\mu$ and $\beta$ we must consider what are the most important features when calculating peak height. From Eqn.~(\ref{eqn:peak_height}) it can be seen that peak height is effectively a convolution between $P(k)$ and $k^2 W^2(kR)$. For both a smooth $k$-space and a STH window function $k^2 W^2(kR)$ exhibits a clear maximum at a specific scale,\footnote{For the smooth $k$-space window function this is only strictly true for $\beta>1/3.12$.} where the associated scale is at $(kR)_{\rm{max}} \equiv \kappa$. This can clearly be seen in the inset panel of Fig.~\ref{fig:window_func}, where we have plotted $(kR)^2 W^2(kR)$ for a few different choices of window functions and parameters. Hence, to first order peak height is set by the amplitude of the linear power spectrum at the associated $k$-scale:
\begin{equation} \label{eqn:nu_prop_P}
\nu^2 \propto R^{3}/P(k_0) \ \ \ .
\end{equation}
where
\begin{equation} \label{eqn:k-scale}
    k_0=\frac{\kappa}{R} \ \ \ .
\end{equation}
Here $R$ is the Lagrangian radius (see Eqn.~(\ref{eqn:lagrangian_rad})) and $\kappa$ corresponds to where $(kR)^2 W^2(kR)$ is a maximum. $\kappa$ is a dimensionless quantity and depends on the choice of window function and associated parameters. For a STH window function, $\kappa=2.08$. For the smooth $k$-space window function, $\kappa$ depends on both $\mu$ and $\beta$ and is found to be
\begin{equation} \label{eqn:const_kappa}
    \kappa = \frac{2.50}{\mu} \bigg ( \frac{1}{3.12 \beta -1} \bigg ) ^{1/3.12 \beta} \ \ \ .
\end{equation}

If we consider contours of constant $\kappa$, we see this equation provides a relationship between $\mu$ and $\beta$ that has the general behaviour of the observed degeneracy, a roughly $\mu=1/\beta$ behaviour. Indeed, we find that this relationship follows almost perfectly the observed degeneracy for an appropriate choice of $\kappa$. This is shown in Fig.~\ref{fig:smooth-kspace} where we have plotted lines of constant $\kappa$. The values of $\kappa$ have been chosen by eye to approximately follow the optimal $\mu$--$\beta$ relation and correspond to $\kappa=9$ and $2$ for constraining $c$ (left panel) and $\alpha$ (right panel), respectively. In detail, it appears that the degeneracy at $\beta \approx 0.5$ is not completely characterised by this relation. In this region of the parameter space the peak in $(kR)^2 W^2(kR)$ is not as clearly defined and therefore higher-order terms will play a more significant role, implying that the simple approximation of $\nu \propto R^{3}/P(k_0)$ will not be as accurate. 

The above results suggest a rather simple interpretation of what sets the average density profiles of DM haloes. It is, to a good approximation, the amplitude of the linear power spectrum at an associated $k$-scale, given by Eqn.~(\ref{eqn:k-scale}), with the one complication that $c$ and $\alpha$ appear to be set by fluctuations on \textit{different} scales.  The concentration of DM haloes is set by smaller scale fluctuations than the shape parameter, by roughly a factor of $4.5$, with the shape parameter matching closely the same value of $\kappa$ for the standard spherical top hat window function. However, it is not clear why this should be the case, and the physical origin of these two preferences requires further study.

The result that at $\beta \approx 1$ there are optimal choices for the smooth $k$-space window function where $\mu \neq 1$ (for halo concentration at least), as well as the dominant factor not being $\mu$ or $\beta$ directly but rather the resulting value of $\kappa$, implies that a STH-like window function can also lead to universal behaviour if an equivalent parameter to $\mu$ is introduced. Let us generalise the STH window function as follows
\begin{equation}
    \label{eqn:sth_generalised}
    W_{\rm{STH,general}}(kR)=\frac{3}{(\mu_{\rm{g}} kR)^3}[\sin(\mu_{\rm{g}} kR)-\mu_{\rm{g}} kR\cos(\mu_{\rm{g}} kR)] \ \ \ .
\end{equation}
Here the window function is identical to the standard definition (see Eqn.~(\ref{eqn:sth})) but with an additional free parameter, $\mu_{\rm{g}}$, that behaves the same as the parameter $\mu$ for the smooth $k$-space window function. In Fig.~\ref{fig:optimal_gSTH} we allow $\mu_{\rm{g}}$ to vary to minimise \chir, as was done for the smooth $k$-space window function. Unlike the smooth $k$-space window function, the generalised STH does not exhibit any degeneracies and there are clearly defined optimal values for $\mu_g$.  We find that for the concentration the optimal value is $\log \mu_{\rm{g}}=-0.67$ with $\chi^2_r = 2.12$, while for the shape parameter $\log \mu_{\rm{g}}=-0.01$ with $\chi^2_r = 1.83$. The optimal values for \chir~are comparable to those found for the smooth $k$-space filter. The associated values of $\kappa$ are $\kappa=9.73$ and $2.13$ for the concentration and shape parameters, which are again comparable to the values of $\kappa$ that match the observed degeneracy between $\mu$ and $\beta$ for the smooth $k$-space window function. 

\begin{figure*}
    \centering
    \includegraphics[width=2\columnwidth]{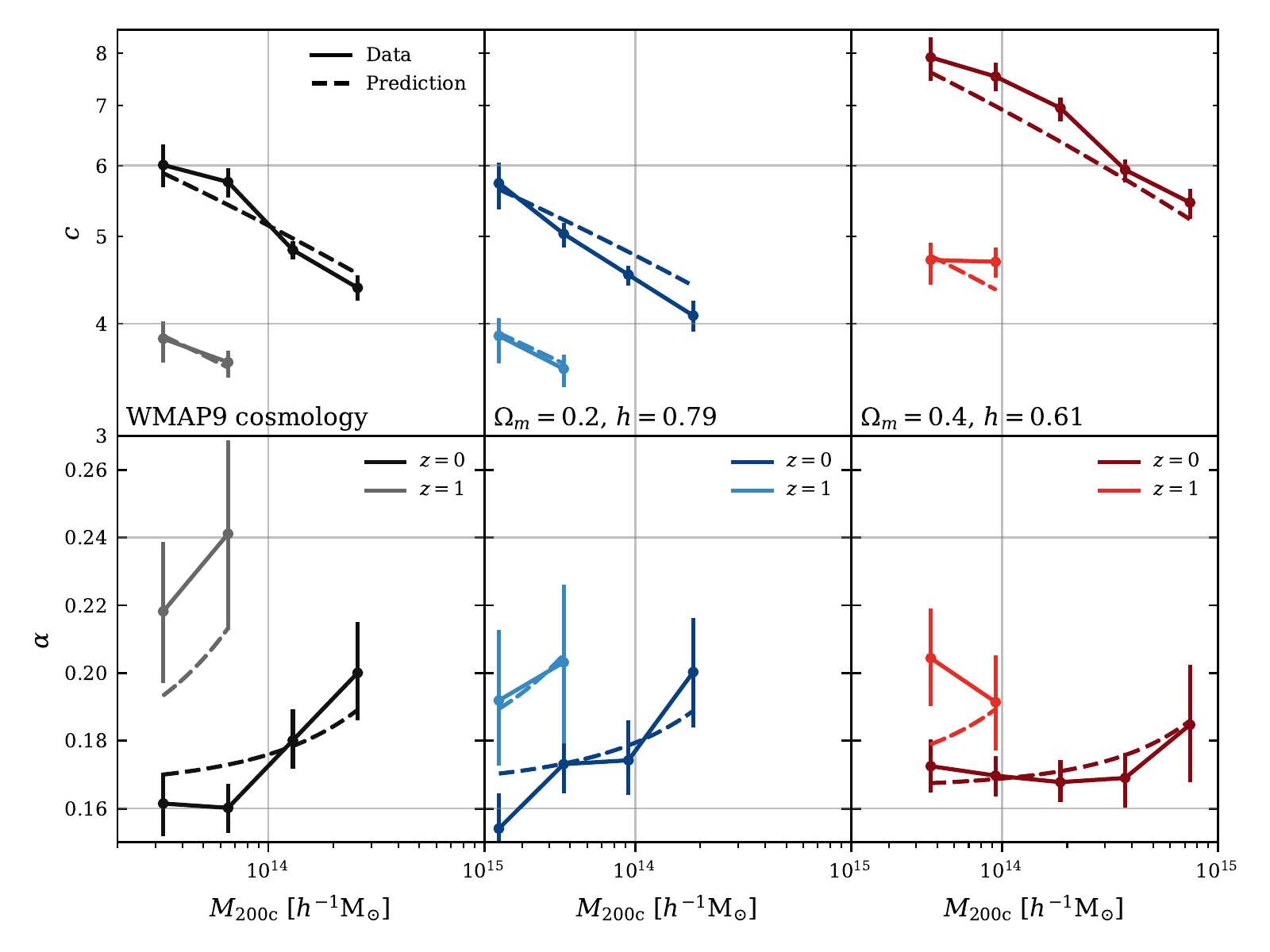}
    \caption{Resulting $c$--$M_{\rm{200c}}$ ({\it top}) and $\alpha$--$M_{\rm{200c}}$ ({\it bottom}) relations for the WMAP 9-yr cosmology ({\it left}), $\Omega_{m}=0.2$ ({\it middle}) and $\Omega_{m}=0.2$ ({\it right}) cosmologies. For each cosmology the relations are shown at $z=0$ and $1$ (see legend). The solid lines with errors represent the data from the simulations while the dashed lines the predictions from the model. In general both $c$ and $\alpha$ are accurately predicted by the model with any differences being within $5\%$ (or approximately one sigma), demonstrating that the model generalises to cosmologies with different background expansions as well as changes to the linear power spectrum.}
    \label{fig:EdS test}
\end{figure*}

In the above discussion, and throughout the paper, we have adopted a single halo mass definition ($M_{\rm{200c}}$) and argued that the two density parameters are effectively set by fluctuations at different physical scales, as described by the optimal window function. However, there is an alternative interpretation that is consistent with the results and formalism presented. As mentioned in Section~\ref{section:peak height definition} the parameter $\mu$, or $\mu_g$, is equivalent to changing the mass associated with the halo. Therefore, an alternative interpretation from the above discussion is to assign a different masses, with a fixed STH window function, for the two density parameters. The shape parameter would therefore use the standard $M_{\rm{200c}}$ definition, while halo concentration would favour a mass definition of $\mu_{g,\rm{c}}^3$=$(10^{-0.67})^3 \approx 0.01$ $M_{\rm{200c}}$, i.e., treating the halo as two orders of magnitude smaller mass. A rough calculation, assuming and Einasto profile with $c=5$ and $\alpha=0.18$, suggests that this would require an overdensity definition of $\Delta \sim 10^5 $, which is significantly larger than most standard mass definitions commonly used. Additionally, this mass definition would represent only a fraction of the amount of accreted matter in virial equilibrium within the halo, and therefore would not represent a physically meaningful quantity. For both these we prefer the interpretation that a single mass definition is used, specifically $M_{\rm{200c}}$, with $\alpha$ and $c$ being set by fluctuations at different associated scales.

\subsection{Predicting the density profile of DM haloes} \label{section:the model}

To develop a model that is able to predict halo concentration and shape parameter for a general cosmology a choice for the best window function must be made. As there are strong degeneracies between $\mu$ and $\beta$ there is no unique choice. We therefore choose to instead use the generalised spherical top hat window function, which we have demonstrated provides equally as universal $c$--$\nu$ and $\alpha$--$\nu$ relations. Using this window function also has the advantage that it reduces the number of free parameters in the model as well as allowing for a more intuitive interpretation of its results, i.e., it corresponds to the density rms averaged over a sphere. For this, we use the optimal parameters found in the previous section, specifically $\log \mu_g=-0.67$ and $-0.01$ for $c$ and $\alpha$. We denote the peak height values calculated with these two choices of window function as \nuc~and \nualpha, respectively.

In Fig.~\ref{fig:optimal_peak_height} we present the resulting $c$--\nuc~and $\alpha$--\nualpha~relations in left and right panels, respectively. In the top row of each plot we show $c$ and $\alpha$ as a function of their respective peak height for all eleven cosmologies studied at $z=0,0.5,1,1.5$ \& $2$. Each data point represents a mass bin from its associated cosmology. The choice of cosmology is specified by the colour, matching that from Figs.~\ref{fig:power_spectra} \& \ref{fig:peak_hight_relation}. The redshift is then specified by the style of the data point (see legend). In general, both $c$ and $\alpha$ are very close to a single function, as expected from the small \chir~values for these choices of window function; \chir$= 1.83$ and $2.12$ for $c$ and $\alpha$, respectively. 

In general most data points lie within $10$\% of the prediction (black dashed line) and the data points that lie significantly further away than this tend to be those with particularly large error bars, but still within a few standard deviations.  It is clear from Fig.~\ref{fig:optimal_peak_height} that using these window functions leads to significantly more universal $c$ and $\alpha$ peak height relations compared with using the standard spherical top hat function in Fig.~\ref{fig:peak_hight_relation}. 

Plotted with black dashed lines in each panel of Fig.~\ref{fig:optimal_peak_height} are the best-fit relations that we use in our model to predict $c$ and $\alpha$. Specifically we use,
\begin{equation} \label{eqn:c-nu}
    c=4.39 \nu_c^{-0.87} \ \ \ ,
\end{equation}
and
\begin{equation} \label{eqn:alpha-nu}
    \alpha=8.52 \times 10^{-4} \, \nu_{\alpha}^4+0.166 \ \ \ .
\end{equation}
Here \nuc~and \nualpha~are calculated using Eqns.~(\ref{eqn:peak_height_1}--\ref{eqn:lagrangian_rad}) and a generalised spherical top hat window function (Eqn.~(\ref{eqn:sth_generalised})). \nuc uses the parameters $\log \mu_g=-0.67$, while \nualpha uses $\log \mu_g=-0.01$.

\begin{table*}
\caption{Optimal \chir~values for different choices of window functions, with the associated optimal parameters. The value of \chir~is calculated by fitting a second order polynomial. For the smooth $k$-space filter there is no unique choice of $\mu$ and $\beta$ that gives a minimum value of \chir, the parameters provided here are just one such possible combination. We have also provided the \chir~values for the prediction of $c$ for two models from the literature for comparison.}
\begin{tabular}{lllll}
\hline
Model or window function                                             & $\chi_{r}^2$~for $c$ & $\chi_{r}^2$~for $\alpha$ & Parameters for $c$     & Parameters for $\alpha$ \\
\hline
Smooth $k$-space (Eqn.~(\ref{eqn:smooth_k}))                    & $2.10$               & $1.83$                    & $\log \mu=-0.64$, $\beta=2$ & $\log \mu=-0.02$, $\beta=2$      \\
Standard spherical top hat (Eqn.~(\ref{eqn:sth}))    & $31.4$               & $1.83$                    & --                     & --                      \\
Generalised spherical top hat (Eqn.~(\ref{eqn:sth_generalised})) & $2.12$               & $1.83$                    & $\log \mu_g=-0.67$            & $\log \mu_g=-0.01$              \\
                                                                     &                      &                           &                        &                         \\
\citet{Diemer_2019}                                                   & 23.8                 & --                        & --                     &        --                 \\
\citet{Ludlow_model}                                                   & 8.30                 & --                        & --                     &        --               \\
\hline
\label{table_chi2_values}
\end{tabular}
\end{table*}

Note that, although a general second order polynomial was used to calculate \chir~values when determining the optimal window function, we re-parameterise these here to better represent the observed trends as well as to have empirical relations that will more reliably extrapolate beyond the values of \nuc~and \nualpha~sampled in this work. For instance, the $c$--$\nu_c$ relation appears to follow very closely a simple power law, allowing the relation to be expressed with only two free parameters. Additionally, using a second order polynomial in log space for the $\alpha$--\nualpha~relation would predict an increase in the value of $\alpha$ as $\nu_{\alpha} \to 0$. There is no indication from this or other work \citep[e.g.][]{Gao_2008,Ludlow_2016} that such an increase would occur, and it seems more likely that $\alpha$ approaches a constant as \nualpha~approaches zero. Such a behaviour is better represented in our chosen parameterisation in Eqn.~(\ref{eqn:alpha-nu}). Using these alternative parameterisations to predict $c$ and $\alpha$ gives consistent \chir~values as found in the previous section using a more complex second order polynomial.

There is much debate in the literature around the form of the $c$--$M$ relation at high masses ($M_{\rm{200c}} \gtrsim 10^{14}$ at $z=0$ for a cosmology close to our own Universe), with some works reporting and upturn in halo concentration at high values of $\nu_{\rm{STH}}$ \citep[e.g.][]{Prada_2012,Diemer_2015} while others see no evidence for an upturn but, in some cases, do report a minimum concentration \citep[e.g.][]{Zhao_09,Ludlow_2014, Correa}. The nature of the high-mass end of the $c$--$M$ relation depends on how the data is processed; if an unbiased sample of haloes is used then there is expected to be an upturn, while if a relaxation cut is applied (as adopted in this work) the preference for an upturn disappears (see \citealt{Ludlow_2012} for more details). In the present study we see no clear evidence of either an upturn or a minimum concentration.\footnote{We have quantitatively verified this by fitting a power law plus a constant (to represent a minimum concentration) and a double power law (to represent an upturn in concentration) to the $c$--\nuc~relations observed in Fig.~\ref{fig:optimal_peak_height}. In both cases a single power law is preferred.} A potential explanation for this is that the inferred values of $c$ depend on whether a free or fixed shape parameter is used, hence the largest discrepancy between our work and those using a fixed shape parameter is expected at high values of \nusth~where $\alpha$ exhibits the strongest mass dependence.   But we note that it is also possible that such features may be present at sufficiently large values of peak height not sampled in this work.

In Table~\ref{table_chi2_values} we present the \chir~values calculated from the halo concentration for few choices of the window function as well as comparing to some models in the literature.  Specifically, the models of \cite{Ludlow_model} and \cite{Diemer_2019}. The publicly available code \texttt{COLOSSUS} \citep{colossus} has been used to generate the quantitative predictions of these two models. We compare to these models as they are designed to predict halo concentration for a general cosmology and were found in \citet{brown2020} to reproduce the general behaviour observed in those simulations. It is clear by the \chir~values shown in Table~\ref{table_chi2_values} that our new model matches more closely the concentrations observed in these simulations.

There are a few key differences between how the concentration of haloes are inferred in our analysis and in these previous studies. Firstly, both these models infer the concentration--mass relation averaged over fits to individual haloes, whereas we have used stacked density profiles. Secondly, they have adopted a fixed shape parameter, $\alpha$, when developing and calibrating their models, as was also done for the concentrations presented in \citet{brown2020}. This was achieved either by explicitly fixing $\alpha$ in the Einasto profile or by using a fitting formula without a comparable shape parameter (i.e., an NFW profile). Allowing both the concentration and shape parameter to be free in the present study, this has arguably led to more accurate measurements of both parameters, which in turn has led to a more accurate model for these quantities. 

\section{Testing the model} \label{Section:EdS test}

In this section we study the predictions of our empirical model for $c$ and $\alpha$ and check that they generalise to cosmologies not already studied here.  One key aspect that remained fixed in the cosmologies used to develop and calibrate the model was the background expansion, with all simulations sharing the same best-fit WMAP 9-yr cosmological parameters: $h=0.7$, $\Omega_{\rm{m}}=0.2793$, $\Omega_{\rm{b}}=0.0463$ and $\Omega_{\Lambda}=0.7207$.  We therefore test the model against two additional cosmologies with distinctly different background expansions. We consider cosmologies with higher and lower matter densities, $\Omega_{\rm{m}}$. Specifically, we study cosmologies with $\Omega_{\rm{m}}=0.2$, $\Omega_{\Lambda}=0.8$, $h=0.79$ and $\Omega_{\rm{m}}=0.4$, $\Omega_{\Lambda}=0.6$, $h=0.61$. Here we have chosen $\Omega_{\rm{m}}$ and then varied $h$ to keep the same distance to the surface of last scattering (which is well-determined from the CMB), we have also enforced that the cosmologies are spatially flat. Additionally, these cosmologies are normalised to the same value of $\sigma_8$, so that there are approximately the same abundance of haloes in the simulations.  We have also kept the ratio of dark matter to baryons, i.e. $\Omega_{\rm{c}}/\Omega_{\rm{b}}$, fixed. The technical details of the simulations are the same as those studied throughout this paper (e.g., a box size of $400h^{-1}$Mpc with $1024^3$ particles, see Section~\ref{section:technical_section} for details).

There are multiple ways in which the different background expansions will affect the evolution and final density profiles of the DM haloes. The most obvious aspect is the redshift evolution of the density fluctuations, as described through the linear growth factor, which will be distinctly different for these cosmologies. This difference will in turn affect the evolution and growth of the internal properties of the DM haloes. However, a more subtle way that changing the background expansion affects both the model and the results is through the mass definition. In this work we have chosen to use a $M_{\rm{200c}}$ mass definition, meaning that the halo mass and radius are defined so that the mean density within $R_{\rm{200c}}$ is $200 \rho_{\rm{crit}}$. Therefore, changing the background expansion not only changes how density fluctuations grow but also the density used to define the mass of the halo, which in turn affects the associated Lagrangian radius and effective scale in the liner power spectrum that sets the peak height value. Testing against these cosmologies will allow us to assess whether both these aspects, the change in growth of the density fluctuation and change in the mass definition, are accurately modelled for a general $H(z)$.

In Fig.~\ref{fig:EdS test} we present the results for these cosmologies, with the associated errors, alongside our predictions for $c$ and $\alpha$. Compared to the fiducial WMAP 9-yr cosmology, the $\Omega_m=0.2$ cosmology matches very closely the mass and redshift evolution while the $\Omega_m=0.4$ one exhibits a much clearer difference, particularly resulting in higher concentrations than the two other cosmologies. It can been seen that our model accurately predicts the mass and redshift evolution for these cosmologies. Most points are well within the errors, with any outlying point being of approximately only one standard deviation away or within $5\%$ of the observed value. It appears that the model and results of this paper therefore do generalise to cosmologies with distinct background expansions. As can be seen in Fig.~\ref{fig:EdS test}, the evolution of $c$ and $\alpha$ as a function of mass and redshift for multiple cosmologies is rather complex. However, this complexity is naturally explained as a single dependence on \nuc~and \nualpha, as demonstrated by the accuracy of the model.

Although the changes studied in this work demonstrate significant differences to the underlying cosmology, both through the linear power spectrum and the background expansion, we have not tested it for even more extreme variations than presented here. The accuracy of the model may be reduced in these regimes, particularly for significantly larger or smaller peak height values than sampled by these simulations. For example, in a cosmology with a truncated power spectra (typically associated with warm dark matter), the $c$-$M$ relation is not expected to be monotonic but instead exhibit a maximum concentration \citep[e.g.][]{Ludlow_model}. For such a cosmology, \nuc~ would tend to a constant at small masses. Hence our model, with a single relation between $c$ and \nuc, would not fully capture the expected turnover.

\section{Summary and conclusions} \label{Section:summary}

The aim of this work has been to create a model that links the fluctuations in the initial linear power spectrum with the resulting density profile of DM haloes, modelling the dependence as a function of mass, redshift and cosmology. To fully describe the density profiles observed in cosmological simulations two parameters are required, halo concentration, $c$, and the shape parameter, $\alpha$. We therefore aimed to create a model that consistently predicts both $c$ and $\alpha$ in a consistent and physically-motivated framework. To this end, we have studied how $c$ and $\alpha$ vary as a function of peak height, $\nu$, a quantity previously shown to correlate strongly, though not perfectly, with both $c$ and $\alpha$ and which is used in the Press-Schechter formalism (see Section~\ref{section:peak height definition} for definitions). We have explored free aspects of the formalism, focusing particularly on the window function, to determine if the relation between both $c$ and $\alpha$ and peak height can be made to be universal, i.e. are a single function for all cosmologies and redshifts. The results of our work can be summarised as follows:

(i) In this work we have used a subset of the cosmological simulations first presented in \citet{brown2020} to study the cosmological dependence of the density profile of dark matter haloes, specifically using the `Planck pivot' and `$k_{\rm{pivot}}=1 h$ Mpc$^{-1}$' suites. For these simulations the slope and amplitude of the initial linear power spectrum has been systematically varied, resulting in haloes with a diverse range of formation and evolution histories. In Section~\ref{section:technical_section} we present the details of the simulations and how the data has been processed to obtain robust and reliable estimates for $c$ and $\alpha$.

(ii) To explore a wide range of possible window functions we used a versatile functional form known as the smooth $k$-space window function (Eqn.~(\ref{eqn:smooth_k}); see also \citealt{Leo_2018}), which is introduced and discussed in Section~\ref{section:window function definitions} (see Fig.~\ref{fig:window_func}). There are two free parameters associated with the smooth $k$-space window function: $\mu$ and $\beta$. $\mu$ determines the effective scale of the transition from unity to zero in the window function while $\beta$ controls how quickly this transition occurs. 

(iii) To quantify how close to universal the $c$--$\nu$ (or $\alpha$--$\nu$) relation is, we fitted a second order polynomial and evaluated the $\chi^2$ error (quoting the reduced $\chi^2$ value throughout) for the given relation, see Section~\ref{section:determining universality}. We studied how \chir~varied as a function of $\mu$ and $\beta$ (see Fig.~\ref{fig:smooth-kspace} in Section~\ref{section:optimal window func}). It was found that there are indeed choices of $\mu$ and $\beta$ that result in universal $c$--$\nu$ and $\alpha$--$\nu$ relations with minimal values of \chir$=2.10$ and \chir$=1.83$ for $c$ and $\alpha$, respectively. 

(iv) It was observed that there is a strong degeneracy between $\mu$ and $\beta$ (again, see Fig.~\ref{fig:optimal_peak_height}) with multiple values providing similarly optimal values of \chir. It was found that the dominant factor in setting the peak height is the scale where the window function is a maximum, when plotted as $k^2W^2(kR)$. Therefore, to first order, the peak height is set by the amplitude of the linear power spectrum at the associated $k$-scale described by $\nu \propto R^3/P(\kappa /R)$. $\kappa$ is where the window function (specifically $k^2W^2(kR)$) is a maximum and depends on the given window function, see Section~\ref{section:optimal window func} and Eqns.(\ref{eqn:nu_prop_P}--\ref{eqn:k-scale}). For the smooth $k$-space window function $\kappa$ depends on both $\mu$ and $\beta$, with contours of constant $\kappa$ matching closely the observed degeneracy. 

(v) The optimal window functions, and associated values of $\kappa$, are different for $c$ and $\alpha$. This strongly suggests that these two quantities are set by fluctuations on \textit{different} physical scales in the linear power spectrum.  The optimal values are $\kappa=8.85$ and $\kappa=2.0$ for $c$ and $\alpha$. For $\alpha$ the optimal window function (and value of $\kappa$) match very closely the standard spherical top hat (STH) window function, while for $c$ the optimal values corresponds to smaller scales.  In particular, our analysis indicates that the concentration of haloes is set by fluctuations on scales $\approx 4.5$ times smaller than those that set $\alpha$, or $\approx 1 \%$ of the halo mass.  As an example, for a WMAP-9 yr best-fit cosmology for a halo with mass $M_{\rm{200c}}=10^{13} h^{-1}$M$_{\odot}$, the concentration is set by fluctuations in the linear power spectrum at a scale of $k \approx 3.1$~$h$Mpc$^{-1}$, while the shape parameter is set by fluctuations at $k \approx 0.7$~$h$Mpc$^{-1}$.

(vi) As the relations between peak height and the density parameters can be made to be approximately universal, we are able to create a simple model where $c$ and $\alpha$ depend only on peak height, with the appropriate choice of window function.  Specifically, we introduced a generalised spherical top hat window function (Eqn.~(\ref{eqn:sth_generalised})) with the optimal parameters $\log \mu_g=-0.67$ and $\log \mu_g=-0.01$ for $c$ and $\alpha$ respectively, see Section~\ref{section:the model}. The values for $c$ and $\alpha$ can then be predicted by empirical relations, given in Eqn.~(\ref{eqn:c-nu}) \& (\ref{eqn:alpha-nu}).  The smooth $k$-space window function also produces similarly accurate relations, the only disadvantage being that it requires two free parameters which are strongly degenerate.

(vii) In Section~\ref{Section:EdS test} we tested the reliability and accuracy of our model. When determining the optimal window function all cosmologies used shared the same background expansion histories, but with systemically varied initial linear power spectra.  As such, we chose to test the predictions of our model against two cosmologies with a higher and lower matter density, resulting in distinctly different evolutions of the Hubble parameter $H(z)$. It was found that the model closely matches the observed $c$--$M_{\rm{200c}}$ and $\alpha$--$M_{\rm{200c}}$ relations, with an accuracy typically better than $10 \%$.

It is common to attribute the concentration of a halo to it's formation time, with this interpretation offering an explanation for both the average halo mass dependence as well as scatter in concentration of individual haloes\citep[e.g.][]{NFW2, Wechsler_2006, Ludlow_2014}. Initially, this view may seem at odds with the results presented in this work (as we do not discuss formation time), but the two pictures are not incompatible. In our model, we attribute the density of collapsed DM haloes directly to properties of the underlying cosmology (i.e., $P(k)$), quantitatively described through the peak height variables \nuc~and \nualpha. The halo formation time, on the other hand, can be viewed as a mediator between changes to the cosmology and the resulting response of the density profiles of DM haloes. Indeed, it seems likely that the idea of the halo concentration being set by fluctuation on a particular scale in the linear power spectrum is roughly equivalent to it being set by the formation time of the halo. One limitation of our model, as it is presented here, is that it only described the average density profiles at a fixed mass. There is expected to be scatter at fixed mass, something that can be explained by an equivalent scatter in formation time. However, formation time is not a fundamental quantity but rather depends on the given cosmology. As such, any prediction for the density profiles (using halo formation time) will require some theoretical framework to predict halo formation time (such as extended Press-Schechter theory), with its own potential systematics and limitations.

Interestingly, multiple studies that link concentration with halo formation time \citep[for example][]{NFW2, Ludlow_model} independently identify the same mass scale in their accretion history, specifically $ \approx 1 \%$ of their current mass, as being important (see the papers for the detailed definitions of formation time). Similarly, we find that the concentration of haloes is set by the effective spatial scale that is (traditionally) associated with $\approx 1 \%$ of the halo mass.  In our view, it seems unlikely to be a coincidence that both these models pick out similar mass scales as being in some sense `special', though the physical significance of this finding remains to be elucidated.

To accurately predict the density profile of DM haloes both $c$ and $\alpha$ are required. Our model can therefore be used to improve the predictive power of many other cosmological tools and probes.  For example, by incorporating it into predictions from the halo model \citep[e.g.][]{Smith_2003,Mead_2015} to improving the fit to stacked weak lensing maps \citep[e.g.][]{von_der_Linden_2014,Hoekstra_2015,McClinktock_2019}. Having a model that accounts for changes in $\alpha$ is particularly important for galaxy cluster mass scales.  At these masses $\alpha$ has the strongest mass dependence as well as deviating significantly from a value that closely resembles an NFW profile, i.e. the prediction is that $\alpha>0.18$ at cluster masses. 

One interesting application would be to use the concentration, or shape parameter, mass relations inferred from observations along with the predictions of ours (or similar) models to constrain the underlying cosmological parameters. Although baryonic changes are expected to play a non-negligible role in setting the total (DM and baryons) density and masses of haloes, these effects are much smaller on the DM component. Therefore, these issues can be mitigated by fitting to the DM \textit{only} component in galaxies/clusters and comparing the inferred mass profiles from a DM only simulation, as discussed, e.g., in \cite{Debackere_2021}. Fitting for both halo concentration alongside cluster abundances is a promising way to help further constrain the cosmology of our Universe, as well as identifying potential systematics (as both should infer the same cosmological parameters).

Our work demonstrates the link between the linear power spectra and the extremely non-linear formation and evolution of the internal density profiles of DM haloes. We have demonstrated that there is a clear universality that exists in the density of haloes in cosmologies dominated by collisionless DM, offering deeper insights into the origin of the structure of our own Universe. This universality leads to robust predictions for the density of DM haloes for a wide range of cosmologies that can be in turn used to further constrain the underlying cosmology of our own Universe. 

Finally, we present a publicly available \texttt{Python} module to calculate the predictions of our model for $c$ and $\alpha$ called \texttt{CASPER} (Concentration And Shape Parameter Estimation Routine). All relevant information about installation and usage can be found at \url{https://github.com/Shaun-T-Brown/CASPER}.

\section*{Acknowledgements}
The authors thank the referee, Aaron Ludlow, and Benedikt Diemer for providing invaluable feedback. STB acknowledges an STFC doctoral studentship. This project has received funding from the European Research Council (ERC) under the European Union's Horizon 2020 research and innovation programme (grant agreement No 769130).
This work used the DiRAC@Durham facility managed by the Institute for Computational Cosmology on behalf of the STFC DiRAC HPC Facility. The equipment was funded by BEIS capital funding via STFC capital grants ST/P002293/1, ST/R002371/1 and ST/S002502/1, Durham University and STFC operations grant ST/R000832/1. DiRAC is part of the National e-Infrastructure.

\section*{Data Availability}

Simulations and data available upon a reasonable request to the authors.



\bibliographystyle{mnras}
\bibliography{references} 




\appendix

\section{Resolution and box size study} \label{section:res_study}

\begin{figure*}
    \centering
    \includegraphics[width=\columnwidth]{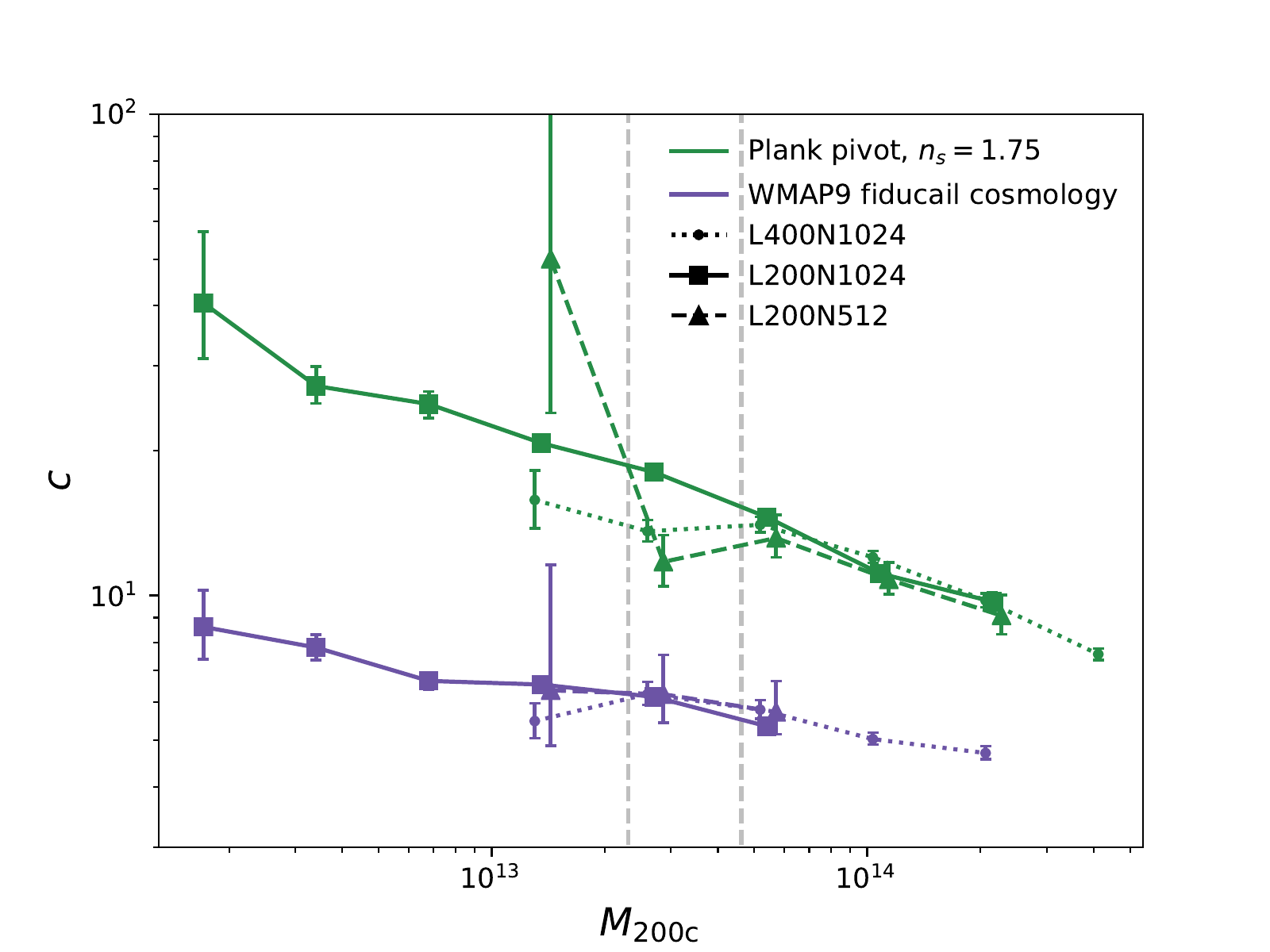}
    \includegraphics[width=\columnwidth]{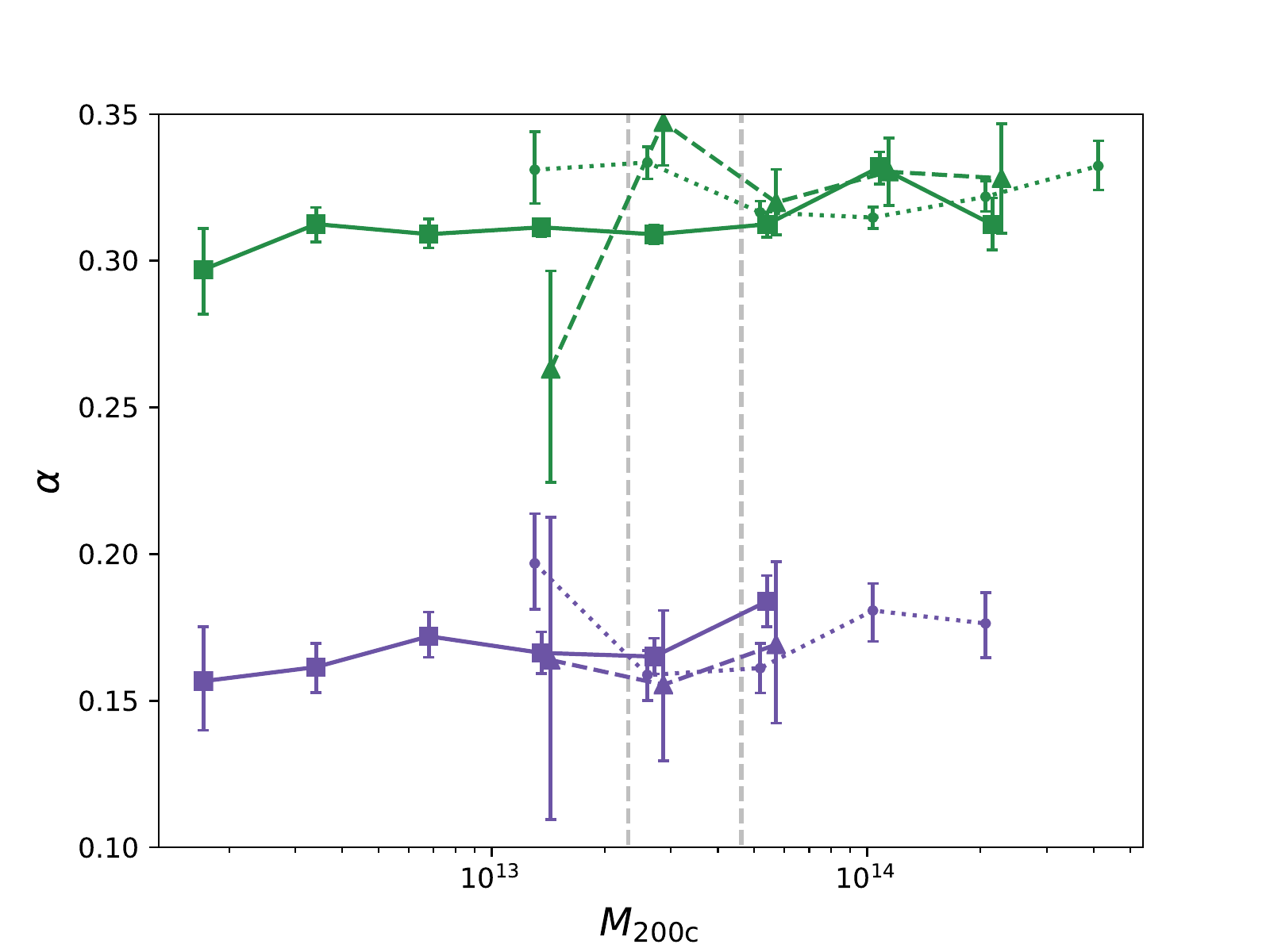}
    
    \caption{The $c$--mass and $\alpha$--mass relations for a variety of box sizes and resolutions to test the convergence and robustness of our results. Presented here are the $z=0$ results for the fiducial best-fit WMAP y-yr cosmology (purple lines) and one that adopts a Planck pivot point and $n_s=1.75$ (green lines), which is the most extreme cosmology studied in this work. The line styles represent the simulation box size and resolution (see legend). Data points have been artificially shifted horizontally for clarity: the L400N1024 data is at the true mass with the L200N1024 and L200N512 multiplied by a arbitrary constant of $1.05$ and $1.05^2$, respectively.  Additionally, the values of $\alpha$ for the Planck pivot cosmology have been increased by a constant of 0.15 with respect to their true value. The values of $c$ have been unchanged due to the data naturally stratifying. The two vertical dashed lines represent haloes with $5,000$ and $10,000$ particles for the L400N1024 and L200N512 simulations.}
    \label{fig:res_test}
\end{figure*}

In this section we present the $c$--mass and $\alpha$--mass relations as a function of varying box size and mass resolution for a few of the cosmologies presented in this paper. We will use the following notation to specify box size and number of particles used in the simulations: L<Boxsize>N<particle number>. For instance, L400N1024 denotes a simulation using a $400$ $h^{-1}\rm{M_{\odot}}$ with $1024^3$ particles, which is the box size and number of particles used throughout the main part of this work.  Presented here are simulations with L200N512, L400N1024 and L200N1024. The details of the simulations and how they are analysed to determine values for $c$ and $\alpha$ are identical to that described in Section \ref{section:technical_section} with the softening length changed appropriately for the higher-resolution L200N1024 simulation, with this simulation using $2$ $h^{-1}$Mpc as opposed to the $4$ $h^{-1}$Mpc used for the other two simulations. With these three simulations we can test both the effects of box size and mass resolution to make sure that neither systematically affect our results. The L400N1024 and L200N512 simulations have the same mass resolution with a different box size, while the L200N1024 and L200N512 simulations share the same box size but have different mass resolutions.

The $c$--mass and $\alpha$--mass relations are presented in Fig.~\ref{fig:res_test} for the three different combination of box size and resolution for the standard WMAP 9-yr best fit and the Planck pivot with $n_s=1.75$ cosmology. As can be seen, both $c$ and $\alpha$ are well converged for all simulations for haloes resolved with an adequate number of particles. It is found that for the most extreme  cosmology we study, i.e. the green lines presented here, at least $10,000$ particles are needed to get sufficiently resolved values for $c$ and $\alpha$. Although not shown here, it is found that only $5,000$ particles are required for all other cosmologies studied.  We therefore use the associated mass cuts when analysing the simulations in this study.

As mentioned the analysis is identical for all simulations. A key part of the analysis is the radial range fit over, which we use $r_{\rm{conv}}<r<0.7R_{\rm{200c}}$, where $r_{\rm{conv}}$ is the convergence radius (see Eqn.~(\ref{eqn:convergence})). $r_{\rm{conv}}$ primarily depends on the number of particles that the halo is resolved with, meaning that the higher resolution simulation (L200N1024) is fit over a wider effective range for the same mass halo. We do not find any systematic difference with mass resolution demonstrating that the $c$ and $\alpha$ are robust to the radial range fit over, as long as an appropriately conservative convergence criterion is used to avoid fitting to the numerical core present. We also do not observe any degeneracy between $c$ and $\alpha$ that is correlated with the radial range being fit over as found in other works \citep[e.g.][]{Udrescu}. We attribute this primarily to fitting to stacked density profiles, resulting in smooth profiles without any discernible features from sub-structure.


\bsp	
\label{lastpage}
\end{document}